\documentclass[journal,10pt]{IEEEtran}

\usepackage{cite}
\usepackage{graphicx}
\usepackage[cmex10]{amsmath}
\usepackage{amsthm}
\usepackage{amsfonts}
\usepackage{caption}
\usepackage{epstopdf}
\usepackage{amssymb}
\usepackage{float}
\usepackage[export]{adjustbox}

\usepackage{color,soul}
\usepackage{subfig}
\usepackage{csquotes}
\usepackage{enumitem}
\usepackage{adjustbox,lipsum}
\usepackage{tabularx,booktabs}
\usepackage{array}
\newcolumntype{L}[1]{>{\raggedright\let\newline\\\arraybackslash\hspace{0pt}}m{#1}}
\newcolumntype{C}[1]{>{\centering\let\newline\\\arraybackslash\hspace{0pt}}m{#1}}
\newcolumntype{R}[1]{>{\raggedleft\let\newline\\\arraybackslash\hspace{0pt}}m{#1}}

\begin{document}
\title{An Intra-Cluster Model with Diffuse Scattering for mmWave Communications: RT-ICM}
\author{Yavuz Yaman, \IEEEmembership{Student Member, IEEE}, and Predrag Spasojevic, \IEEEmembership{Senior Member, IEEE}
\thanks{The authors are with Department of Electrical and Computer Engineering, Rutgers University, Piscataway, NJ 08854 USA (e-mail:yavuz.yaman@rutgers.edu; spasojev@winlab.rutgers.edu).}}

\maketitle

\begin{abstract}
\boldmath
 In millimeter-wave (mmWave) channels, to overcome the high path loss, beamforming is required. Hence, the spatial representation of the channel is essential. Further, for accurate beam alignment and minimizing the outages, inter-beam interferences, etc., cluster-level spatial modeling is also necessary. Since, statistical channel models fail to reproduce the intra-cluster parameters due to the site-specific nature of the mmWave channel, in this paper, we propose a ray tracing intra-cluster model (RT-ICM) for mmWave channels. The model considers only the first-order reflection; thereby reducing the computation load while capturing most of the energy in a large number of important cases. The model accounts for diffuse scattering as it contributes significantly to the received power. Finally, since the clusters are spatially well-separated due to the sparsity of first order reflectors, we generalize the intra-cluster model to the mmWave channel model via replication. Since narrow beamwidth increases the number of single-order clusters, we show that the proposed model suits well to MIMO and massive MIMO applications. We illustrate that the model gives matching results with published measurements made in a classroom at 60 GHz. For this specific implementation, while the maximum cluster angle of arrival (AoA) error is $1$ degree, mean angle spread error is $9$ degrees. The RMS error for the cluster peak power is found to be $2.2$ dB.
\end{abstract}

\section{Introduction}\label{intro}

\IEEEPARstart{M}{illimeter}-wave (mm-Wave) communication is continuing to emerge with several advantages over the current wireless bands such as higher throughput, lower latency, reduced interference, and increasing network coordination ability. Many indoor and outdoor measurements aim at modeling the mmWave channel characteristics \cite{Xu, Maltsev_office, Maltsev_conference, Murdock, Samimi, Rappaport_aoa, Thomas, Rangan, Akdeniz}. Beamforming  is typically introduced to compensate for the higher path loss at higher mmWave frequencies \cite{Yaman, Ayach}. However, spatial filtering of the channel requires detailed knowledge of the angle spectrum of the channel. Fortunately, clusters are spatially well-separated in mmWave channels, which allows creating a beam for each cluster \cite{11ay}. Furthermore, as the first order reflections and the direct path cover as high as 99.5\% of the received power \cite{Gentile}, received clusters in mmWave can usually be considered for the first-order reflections only. 
On the other hand, well-known channel propagation mechanisms affect mmWave channels differently. For example, while diffraction contributes to the received power for microwave channels, its contribution is negligible in mmWave channels \cite{Murdock}. Also, scattering is limited in lower frequencies, %and only specific structures such as lampposts, trees, and traffic signs act as scatterers \cite{Rappaport_book} 
whereas, in mmWave channels, even a typical wall can scatter the incoming signal significantly due to the tiny variations on its surface. Hence, a mmWave channel model should take the diffuse scattering into account in order to properly replicate the channel characteristics \cite{Maltsev_office, Gustafson, Pascual3}.
Measurement results show that in some NLOS cases wider beamwidth antennas result in higher received SNR \cite{Rajagopal,Rappaport_aoa} which leads to the fact that array design that would create optimum beamwidth is directly related to the spatial representation of the received cluster. Then, although the clusters can be easily identified, an accurate intra-cluster angular model has vital importance. Hence, knowledge of the detailed cluster angular spectrum is essential for at least two important applications including accurate beam alignment along with an optimum beamwidth and to minimization of inter-beam interference.

Measurements confirm that the mmWave channels are site-specific\cite{Murdock, Smulders} and that the channel characteristics depend highly on the environment \cite{Gustafson_journal}. Hence, creating generic statistical models for typical environments as in the case of microwave bands is difficult \cite{3GPP}. For this reason, researchers tend to propose statistical channel models for specific environments \cite{Smulders, Gentile, Kyro, Murdock, Samimi}. 
%In the IEEE 802.11ad standard documentation \cite{11ad}, conference room \cite{Maltsev_conference}, living room and office cubicle \cite{Maltsev_office} statistical channel models are proposed for 60 GHz. IEEE 802.11ay \cite{11ay} increases the use case list by adding open area, street canyon and hotel lobby environments. \cite{Gentile} gives a model for a data-center, while \cite{Kyro} introduces one for the hospital area at 60 GHz. Campus environment channel model is studied in \cite{Murdock} at 38 GHz and New York City streets are in \cite{Samimi} at 28 and 73 GHz. %These developments suggest that the site-specific modeling such as ray tracing may be more appropriate for mmWave. 
To give generalized models, a hybrid geometry-based stochastic channel model (GSCM) \cite{Steinbauer} that combines stochastic and deterministic approaches was recently introduced \cite{Samimi,Thomas} and adopted by the mmWave wireless standards such as 3GPP \cite{3GPP}, IEEE 802.11ad \cite{11ad}, IEEE 802.11ay \cite{11ay}, MiWEBA \cite{MiWEBA} and  COST2100 \cite{COST2100}. Although the hybrid method is more accurate than the statistical approach, while generating faster and more generalized results than the deterministic approach, nevertheless it does not provide sufficient intra-cluster angular modeling accuracy necessary for beamforming and inter-cluster interference optimizations. Specifically, in 3GPP Channel Model \cite{3GPP, Thomas}, the intra-cluster %time parameters are modeled statistically based on the combination of ray tracing and measurement. But, 
angular modeling is solely based on measurements and the number of paths within the cluster and their powers are fixed for certain type of environments. On the other hand, 60 GHz indoor standards 802.11ad \cite{11ad}, 802.11ay \cite{11ay} and 802.15.3c \cite{3c} adopt statistical intra-cluster model %where the multipath components arrivals are approximated as Poisson process, 
rooted from the S-V model \cite{Saleh, Spencer} and angular behavior of the rays within the cluster are simply modeled as a %Gaussian in \cite{11ad,11ay} and Laplacian in \cite{3c}. 
random variable. While Quasi-Deterministic (Q-D) Channel Model \cite{Maltsev_QD,Weiler} takes mmWave scattering into account, its effect on the spatial domain in the cluster level is not addressed. As a result, to the authors' knowledge, a detailed intra-cluster mmWave channel model that studies the power distribution in angle domain has not been introduced yet.

In this paper, we propose a spatial ray-tracing mmWave intra-cluster channel model (RT-ICM) that takes only first-order reflections into account. In our model, we also add the scattering effect based on the material properties. The model outputs the power distribution both in angle and time domain within the cluster and can be used for both indoor and outdoor mmWave systems in any type of environments given the conditions that the required physical parameters for ray-tracing are provided. We further provide a MIMO channel model that consists of nonoverlapping clusters and discuss that pencil-shape beamwidth provided by massive MIMO allow an increased number of single-order clusters in mmWave. Furthermore, we give the insights that, with the combination of massive MIMO and the proposed channel model, maximum spatial usage of the channel can be achieved using several beams with different beamwidths directed to detected clusters. The advantage of the proposed model is that it provides the accuracy of the deterministic approach and the simplicity of the stochastic approach while comes with an intra-cluster model addition to the hybrid approaches. We also show that the results of the proposed model match well with the published indoor mmWave measurements.

The paper is organized as follows. In Section \ref{sub_modelCluster}, the definition of the cluster used in the model is given along with the assumptions. We introduce a basic geometrical model for a single-order reflection cluster in Section \ref{sub_BGM} and using this model, we propose the intra-cluster model with all aspects in Section \ref{sub_IntraClusterModel}. In Section \ref{Sec_MIMO}, the model is extended to MIMO scenarios and its implementation to the already provided measurements is given in Section \ref{Sec_implementation}. Finally, Section \ref{Sec_Conclusion} concludes the paper.

	\section{Cluster Definition of the Model} \label{sub_modelCluster}
	
\begin{figure}[t]
\centering
\includegraphics[scale=0.5]{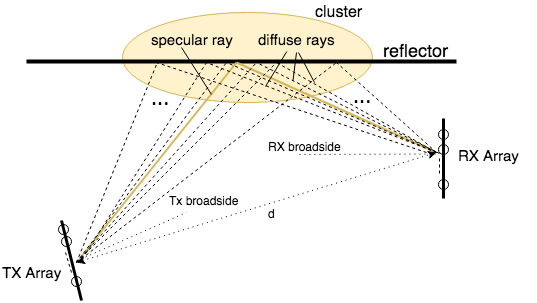} \\
\caption{Cluster definition of the model}\label{clusterdef}
\end{figure}

Let two stationary devices positioned and communicate with each other at a distance as seen from the top view in Fig. \ref{clusterdef}. A rough surface acts as a reflector that creates the cluster which includes several rays that are generated by diffuse scattering. Note that only one ray obeys Snell's Law \cite{Glassner} in the cluster model and it is called \textit{specular ray}, referring to the specular reflection. The others are going to be called \textit{diffuse rays}, referring to the diffuse reflections. Both ray types are shown in Fig. \ref{clusterdef}. Only 2-D azimuthal plane is considered in the model.

In our model, we strict the clusters to be generated only via first-order reflections for NLOS scenarios. We also let each reflector can create only one cluster. In other words, the number of first-order reflectors in the environment equals the number of clusters. Also, when we discuss multi-cluster scenarios, we assume the clusters do not overlap spatially. 

Although we pictured the devices as phased array antennas, the proposed model assumes them as point sources. Hence, each ray leaves from the transmitter, reflects from a unique point at the reflector and then reaches to the receiver with a unique AoA. %The width of the cluster ellipse (the distance between leftmost reflection and rightmost reflection points at the wall surface) is determined by geometry, transmitter beamwidth and the reflector length. The details are given in the next subsection.

	\section{Basic Geometric Model (BGM)} \label{sub_BGM}
	
%In order to output the spatial behavior of the channel, geometrical parameters of the communication environment have to be specified. 
%In this subsection, a geometric simple model of the environment is provided. 
Unless otherwise is stated, all distances are in meters, and the angles are in degrees. We illustrate angles in diagrams with clockwise rows if they are positive, counterclockwise rows otherwise.

		\subsection{Environment Setup} \label{subsub_Environ}
		
\begin{figure}[t]
\centering
\includegraphics[scale=0.55]{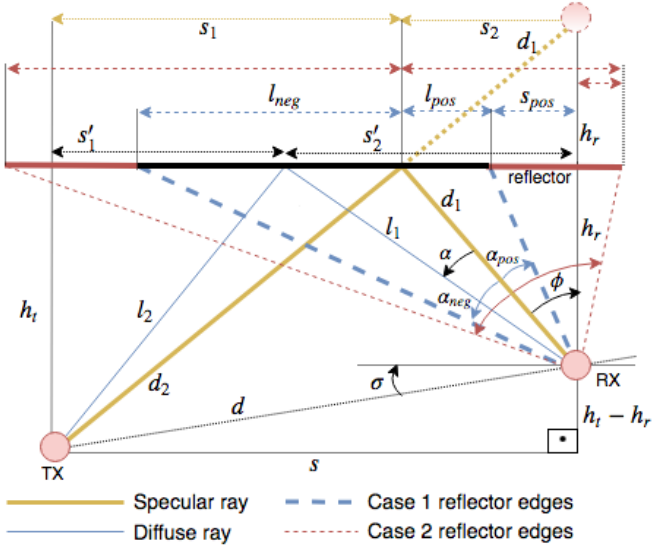} \\
\caption{Basic geometrical model of the cluster}\label{basicmodel}
\end{figure}

Considering the specular ray and one diffuse ray, all main parameters can be defined as seen in Fig. \ref{basicmodel}. As seen, the shortest ray within the cluster is the specular ray. %This can be proved by simply mirroring either the receiver or the transmitter with respect to the wall. 

\begin{table}[!t]
\centering
\caption{Geometrical Notations} 
\label{geom}
\begin{tabular}{|c|l|}
\hline
Notation  & Definition
  \\ \hline \hline
$d$ & length between the transmitter and receiver
\\ \hline
$h_t$ & length between the transmitter and the reflector
\\ \hline
$h_r$ & length between the receiver and the reflector
\\ \hline
$s$ & length between the reflector normal at transmitter (RNT) \\ & and the reflector normal at receiver (RNR)
\\ \hline
$d_1$ & length between specular ray reflection point and the receiver
\\ \hline
$d_2$ & length between specular ray reflection point and the transmitter
\\ \hline
$l_1$ & length between diffuse ray reflection point and the receiver
\\ \hline
$l_2$ & length between diffuse ray reflection point and the transmitter
\\ \hline
$s_1$ & length between specular ray reflection point and the RNT
\\ \hline
$s_2$ & length between specular ray reflection point and the RNR
\\ \hline
$s_1^\prime$ & length between diffuse ray reflection point and the RNT
\\ \hline
$s_2^\prime$ & length between diffuse ray reflection point and the RNR
\\ \hline
$\phi$ & specular ray angle of arrival (AoA) with respect to the RNR
\\ \hline
$\alpha$ & offset AoA between specular and diffuse ray
\\ \hline
$\l_{neg}$ & reflector length that covers diffuse rays with negative $\alpha$
\\ \hline
$\l_{pos}$ & reflector length that covers diffuse rays with positive $\alpha$
\\ \hline
$s_{pos}$ & length between the receiver-side reflector endpoint and the RNR
\\ \hline
$\alpha_{pos}$ & positive offset AoA limit due to reflector length 
\\ \hline
$\alpha_{neg}$ & negative offset AoA limit due to reflector length
\\ \hline
$\Theta$ & beamwidth of the transmitter beam
\\ \hline 
$l_t$ & length of the transmitter side illumination at reflection line
 \\ \hline
$l_r$ & length of the receiver side illumination at reflection line
\\ \hline
$s_t$ & length between the transmitter side beam edge and the RNT
\\ \hline
\end{tabular}
\end{table}

The distances given in the diagram are defined in Table \ref{geom}. Hence, the length of the specular ray is $l_{sp}=d_1+d_2$. Similarly, the length of the diffuse ray is
\begin{equation}
l_{dif}=l_1+l_2 \label{ldif}
\end{equation}

%Finally, $\beta=\phi-\alpha$ is the \textit{diffuse ray AoA with respect to the reflector normal}. %In the figure, it is demonstrated as a positive angle, but it can have negative values too depending on the reflection point of the diffuse ray, existing either from the left side or the right side of the specular ray reflection point. 
The formulation will be setup according to Fig. \ref{basicmodel} and then the validation of the formulas for other scenarios will be checked in Section \ref{subsub_formulValid}. Geometrical derivations are given in Appendix \ref{appen_calculate}.

		\subsection{Calculating Path Lengths and Angles} \label{subsub_Calc}
		
%The goal of the BGM is to find the diffuse ray length for a given $\alpha$ and generalize it for any case of scenarios. 
With the given distances $d$, $h_t$ and $h_r$, specular ray length is $\l_{sp}=\sqrt{s^2+(h_t+h_r)^2}$ where $s=\sqrt{d^2-|h_t-h_r|^2}$.
%Note that the absolute value is inserted to account the case where $h_r>h_t$ which is also shown in Figure \ref{othercase}. 
%\begin{figure}[t]
%\centering
%\includegraphics[scale=0.55]{othercase.png} \\
%\caption{The diagram of the case where the transmitter is closer to the reflector}\label{othercase}
%\end{figure}
%Proceeding with Figure \ref{basicmodel}, 
The specular ray AoA is $\phi=\cos^{-1} \left( (h_t+h_r)/l_{sp} \right)$.
Finally, the diffuse ray length for a given $\alpha$ is  
\begin{equation}
\begin{split}
l_{dif}=&\frac{h_r}{\cos (\phi-\alpha)} \\
&+\sqrt{h_t^2+ \left( s-\frac{h_r}{\cos (\phi-\alpha)}\sin (\phi-\alpha) \right) ^2}
\end{split} \label{ldif_main}
\end{equation}
%Note that for $\alpha=0$, $l_{dif}=l_{sp}$. Intermediate steps are given in Appendix \ref{appen_calculate}.
\begin{figure}[t]
\centering
\includegraphics[scale=0.11]{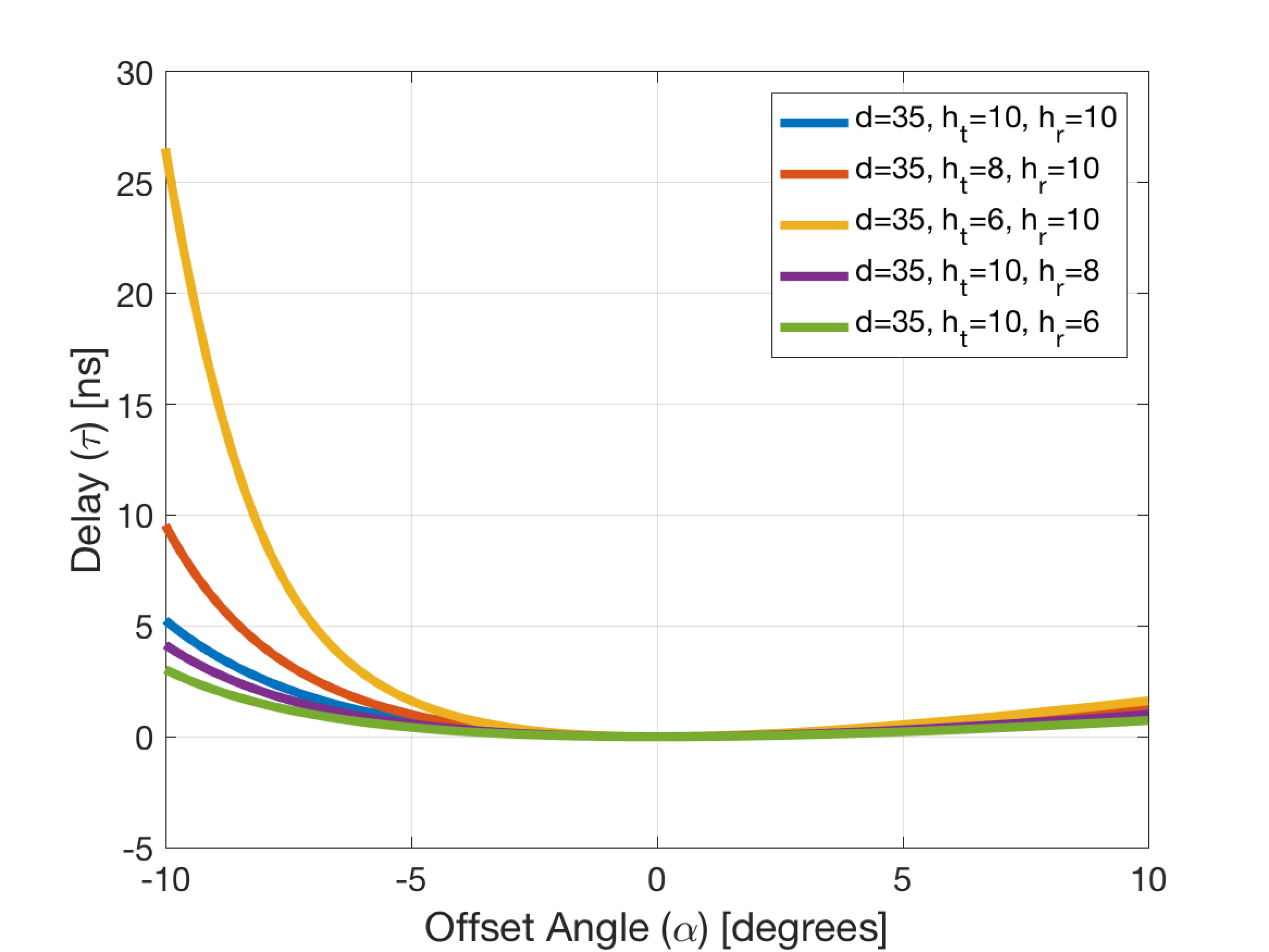} \\
\caption{Relation between offset AoA and diffuse ray delay}\label{delayangle}
\end{figure}

		\subsection{Timing Parameters and Time-Angle Relation} \label{subsub_timeangle}
		
%We seek to express spatial and temporal characteristics of the channel in terms of each other. 
The time of arrival (ToA) of the line-of-sight (LOS) ray is,  $t_{los}=d/c$ where $c$ is the speed of light. %Let the receiver starts receiving at $t=0$, then $t_{los}$ is called the \textbf{time of arrival} of the LOS ray. 
Similarly, $t_{sp}=l_{sp}/c$ and $t_{dif}=l_{dif}/c$ are the ToA of the specular and diffuse ray, respectively. %Hence, from Eq. (\ref{ldif_main}),
%\begin{equation}
%t_{dif}=\frac{\frac{h_r}{\cos (\phi-\alpha)}+\sqrt{h_t^2+ \left( s-\frac{h_r}{\cos (\phi-\alpha)}\sin (\phi-\alpha) \right) ^2}}{c} \label{tdif_alpha}
%\end{equation}
%Finally,
%expresses the time of arrival of the diffuse ray in terms of the offset AoA. 
%Plugging diffuse ray AoA in Eq. (\ref{beta}),
%\begin{equation}
%t_{dif}=\frac{\frac{h_r}{\cos (\beta)}+\sqrt{h_t^2+ \left( s-\frac{h_r}{\cos (\beta)}\sin (\beta) \right) ^2}}{c}\label{tdif_beta}
%\end{equation}
%gives the relation between diffuse ray AoA and the time of arrival.
%\begin{figure}[t]
%\centering
%\includegraphics[scale=0.60]{delayanglefull.eps} \\
%\caption{Delay-offset AoA relation for full range of offset AoA}\label{delayanglefull}
%\end{figure}
%Most of the time, in order to compare the different scenarios, timing characteristics of the channel is defined in terms of delays with respect to a fixed reference ray time of arrival. Hence, 
%As we try to analyze the intra-cluster structure, letting $t_{sp}$ be the reference time,
Finally, $\tau_{dif}=t_{dif}-t_{sp}$ is the diffuse ray \textit{delay} with respect to the specular ray. 

%Eq. (\ref{tdif_alpha}) and (\ref{taudif}) convert the given angular information of the ray to time domain parameters. Note that the function also depends on the geometry of the reflection. 
Fig. \ref{delayangle} displays the $\tau$-$\alpha$ relation for different values of $h_t$ and $h_r$ while $d$ is fixed to $35$ meters. The range for the $\alpha$ is selected as $[-10^{\circ},10^{\circ}]$ which is a typical angle spread of a cluster and the delays are given in nanoseconds. As seen, for any $\{h_t,h_r\}$ pair, the function is not symmetric. That means delays are not necessarily equal for two equal opposite signed offset AoAs. %This phenomenon occurs because of the reason that the two rays, reflect from the right and left side of the specular reflection point at the reflector surface and reach receiver with the same absolute valued offset AoAs, have different lengths.  That can be seen in Eq. (\ref{ldif_main}) where the sine and cosine terms yield different results for the same offset AoAs as following:
%\begin{equation*}
%\cos (\phi-\alpha) \neq \cos (\phi+\alpha)
%\end{equation*}
%\begin{equation}
%\sin (\phi-\alpha) \neq \sin (\phi+\alpha)
%\end{equation}
%After all, the ray lengths are calculated according to AoAs, not offset AoAs. And clearly, although they have same offset AoAs, the AoAs with the respect to the wall normal are different. That is, $|\alpha_1|=|\alpha_2|$, but $\beta_1 \neq \beta_2$.
Another important result is that the delay-angle relation highly depends on the environment. Even a very small change in distances yields much different delays for $\alpha<0$.
% \begin{figure}[t]
%\centering
%\includegraphics[scale=0.5]{twocases.png} \\
%\caption{Diagram of physical explanation of geometry limitation (a) negative  limitation (b) positive  limitation}\label{twocases}
%\end{figure}

%Figure \ref{delayanglefull} shows another case of Eq. (\ref{taudif}) with $d$ is now $200$m and $h_t=h_r=50$m where the equation is plotted for the full range of $\alpha$. The meaning of the "full range" here is the range where a reflection is possible. That is, the geometry of the environment doesn't allow a reflection at the outside of the given range.  The limitations for $\alpha$ is discussed further in Section ‎\ref{subsub_supportReg}. 

%According to Figure \ref{delayanglefull}, delay has an exponential increase at both sides of $\alpha$, only the rate is different and, in fact, strictly depends on the $\phi$. 

		\subsection{Support Region} \label{subsub_supportReg}

\begin{figure}[t]
\centering
\includegraphics[scale=0.6]{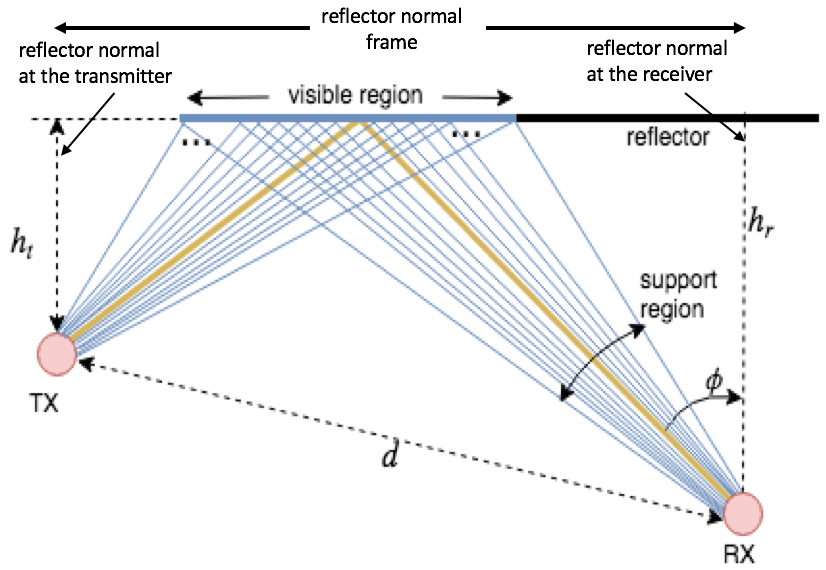} \\
\caption{Diagram of offset AoA limitations}\label{deterministic}
\end{figure}

%So far, we didn't assume any limitation on the environment geometry of the link. In fact, 
Several effects exist in practical scenarios that bound the angle spread of the receiver. We account those constraints on $\alpha$ and call the resultant available range as \textit{support region}. We also define a region on the reflector surface that covers all the reflection points, called \textit{visible region}. The visual meaning of these terms is shown in Fig. \ref{deterministic}. Support region is limited primarily by the reflection geometry, secondarily by the visible region. Visible region is limited by the reflector length and the transmitter beamwidth. We give the ranges in the next subsections for each while the details are provided in Appendix \ref{appen_support}. %In fact, the receiver beamwidth also constraints the offset AoA but we leave its effect to other work to be studied in details, thus we assume the receiver has an omnidirectional antenna pattern for now.

\subsubsection{Reflection Geometry} \label{SSS_limitGeo}

%Although the offset AoA bounds due to the basic geometry are implicitly accounted during the formulation, it is not studied analytically and indeed the limits are not provided. As seen from Figure \ref{delayanglefull}, the delay goes infinity after an offset AoA for each side of the function. This is quite normal as 
%The cosine term in the denominator in Eq. (\ref{tdif_beta}) approaches 0 as $\beta$ gets closer to either $90^{\circ}$ or $-90^{\circ}$. 
%The physical description of the phenomenon is displayed in Figure \ref{twocases} for both sides. Then, 
%\begin{equation*}
%-90^{\circ}<\beta<90^{\circ}
%\end{equation*}
%\begin{equation*}
%-90^{\circ}<\phi-\alpha<90^{\circ}
%\end{equation*}
%\begin{equation}
%\phi+90^{\circ}>\alpha>\phi-90^{\circ} \label{bygeo}
%\end{equation}

%gives the upper and lower limits for offset AoA. 

%Note that we formulate the limits based on the scenario where the reflector is equidistant to the transmitter and the receiver, i.e. $h_t=h_r$. However, %while $d$ doesn't affect the angular parameters of the environment,
$h_t$ and $h_r$ change the geometry drastically as seen from Fig. \ref{delayangle}. Hence, two cases, $h_t>h_r$ and $h_t<h_r$, should be checked separately. % for an accurate calculation of reflection geometry limitation.
% \begin{figure}[t]
%\centering
%\includegraphics[scale=0.5]{case1.png} \\
%\caption{Geometry limitation when $h_t>h_r$ for negative $\alpha$ and positive $\alpha$ }\label{case1}
%\end{figure}

For case $h_t>h_r$, as seen from Fig. \ref{basicmodel}, a positive tilt angle $\sigma$ is introduced that needs to be taken into account and calculated as $\sigma=\sin^{-1}\left( (h_t-h_r)/d \right)$.

Then, accounting the leftmost and rightmost possible reflections, $\phi-\sigma+90^{\circ}>\alpha> \phi-90^{\circ}$. Simliarly, for $h_t<h_r$, the range is given as  $\phi+90^{\circ}>\alpha> \phi-\sigma-90^{\circ}$ as $\sigma<0$ now and bounds the negative $\alpha$. %and the case is illustrated in Figure \ref{case2}.% where $\alpha_1$ and $\alpha_2$ show offset AoAs of the leftmost negative and rightmost reflections, respectively.
% \begin{figure}[t]
%\centering
%\includegraphics[scale=0.5]{case2.png} \\
%\caption{Geometry limitation when $h_t<h_r$ for negative $\alpha$ and positive $\alpha$ }\label{case2}
%\end{figure}

%As a result, the limitation due to the geometry is summarized in Table \ref{geometrylim}. It is obvious that both unequal distance cases support equal case when $\sigma=0$, but we add a row for $h_t=h_r$ for describing the case, explicitly.
%\begin{table}[!t]
%\centering
%\caption{Summary of Geometry Limitation} %\color{blue}{ *** Remove any notation or parameters not used ***}
%\label{geometrylim}
%\begin{tabular}{|c||c|}
%\hline
%Case  & Support Range for $\alpha$
%  \\ \hline \hline
%\begin{tabular}{@{}l@{}}$h_t=h_r$ \end{tabular}   & \ \ \ \ \  $\phi+90^{\circ}>\alpha>\phi-90^{\circ}$  
%\\ \hline
%\begin{tabular}{@{}l@{}}$h_t\geq h_r$ \end{tabular}   & $\phi-\sigma+90^{\circ}>\alpha>\phi-90^{\circ}$ 
%\\ \hline
%\begin{tabular}{@{}l@{}}$h_t<h_r$ \end{tabular}   & \ \ \ \ \ \ \ \ \ \ \ $\phi+90^{\circ}>\alpha>\phi-\sigma-90^{\circ}$  
% \\ \hline                                                                                                           
%\end{tabular}
%\end{table}

\subsubsection{Visible Region} \label{SSS_limitRef}

%Notice that we have not specified a reflector length in the formulation when calculating the ray AoAs and lengths. In fact, so far, we were assuming the length of the reflector is infinity. Obviously, that is impractical and a finite length of a reflector has the effect of limiting the range of the cluster as well as the angle AoAs, hence indirectly the offset AoAs. Thus, a finite length effect has to be taken into account analytically to model the cluster more practically. In this section, we parameterize the reflector length and embed into the support range of the offset AoA.

\begin{figure}[t]
\centering
\includegraphics[scale=0.5]{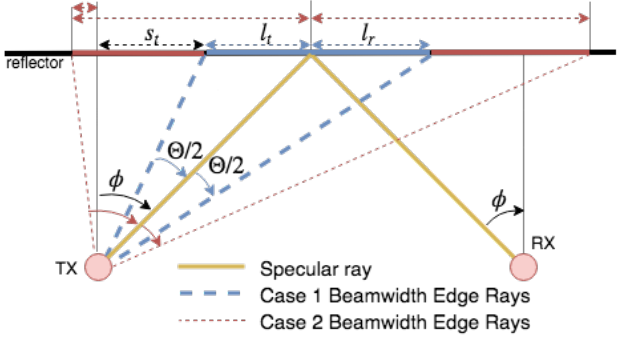} \\
\caption{Transmitter beamwidth limitation}\label{txbw}
\end{figure}

Reflector length determines the visible region geometrically, whereas system parameter transmit beamwidth is another limitation. Reflector length limitation is illustrated in Fig. \ref{basicmodel} for two cases. Ignoring the misalignment problems and sidelobes in the radiation patterns, we consider the transmit beam is steered towards the specular ray and divide it into two to determine the covered region on the reflection line. The diagram in Fig. \ref{txbw} visualizes the approach for two cases. Related parameters are listed in Table \ref{geom}. 
Hence, minimum of two limitations at both sides will determine the \textit{visible region}. Analytically,

\begin{equation*}
w_t=\min (l_t, l_{neg})
\end{equation*}
\begin{equation}
w_r=\min (l_r, l_{pos}) \label{wtwr}
\end{equation}

where $w_t$ and $w_r$ are the transmitter and receiver side visible region lengths, respectively. As an example, in Fig. \ref{deterministic}, we let the $w_t$ is limited by the reflector length. And the $w_r$ is limited by the transmitter beamwidth. Note that the knowledge of the reflector length is not enough as $l_{pos}$ is not necessarily equal to $l_{neg}$. Hence, along with $d$, $h_t$, $h_r$ and $\Theta$, both sides reflector lengths, $l_{pos}$ and $l_{neg}$, should be given as inputs to the model as well. Derivations of $l_t$ and $l_r$ are given in Appendix \ref{appen_support}.

In order to determine a range for the offset AoA due to the visible region, we backtrack the received rays that reflect from the endpoints of the region. Then, the offset AoA upper and lower bounds due to the visible region are
\begin{equation}
\alpha_{neg}^{\prime}<\alpha<\alpha_{pos}{\prime} \label{geo}
\end{equation}

where $\alpha_{neg}{\prime}=\phi-\tan^{-1}\left( (d_1\sin \phi+w_t)/h_r\right)$ and $\alpha_{pos}{\prime}=\phi-\tan^{-1} \left( (d_1\sin \phi-w_r)/h_r \right)$.

%Notice that, since the range limits don't depend on $h_t$, the cases $h_t>h_r$ and $h_t<h_r$ won't affect the support range. However, we should take into account the cases where the wall length exceeds the wall normal at transmitter and receiver. Although some of those cases result in violating physics in intermediate steps (such as some values of lengths yield negative), the formulation doesn't change since the resultant range is accurate for those cases. We will elaborate the possible cases and validation of all formulas for those cases in Section ‎\ref{subsub_formulValid}.

\begin{table}[!t]
\centering
\caption{Resultant Support Range for $\alpha$} 
\label{supportrange}
\begin{tabular}{|c||c|}
\hline
  Case  &  Support Range 
  \\ \hline \hline
$h_t \geq h_r$  & \begin{tabular}{c|l} $\alpha^-$ & $\max \left\lbrace \phi-90^{\circ},\alpha_{neg}{\prime} \right\rbrace $ \\\hline
$\alpha^+$ & $\min \left\lbrace \phi-\sigma+90^{\circ}, \alpha_{pos}{\prime} \right\rbrace$ 
 \end{tabular}
\\ \hline 
$h_t<h_r$ & \begin{tabular}{c|l} $\alpha^-$ & $\max \left\lbrace \phi-\sigma-90^{\circ},\alpha_{neg}{\prime} \right\rbrace $ \\ \hline 
$\alpha^+$ & $\min \left\lbrace \phi+90^{\circ}, \alpha_{pos}{\prime} \right\rbrace$ 
 \end{tabular}
\\  \hline                                                                                                         
\end{tabular}
\end{table}

%\paragraph{Resultant Support Region} \label{SSS_resultSupport}
Finally, combining with the reflection geometry limitation and having the tighter constraint on both sides, we give the expressions for the resultant support region for $h_t \geq h_r$ and $h_t < h_r$ in Table \ref{supportrange} with $\alpha \in [\alpha^-, \alpha^+]$.
		\subsection{Formulation Validation for Other Scenarios} \label{subsub_formulValid}
		
\begin{figure}[t]
\centering
\includegraphics[scale=0.5]{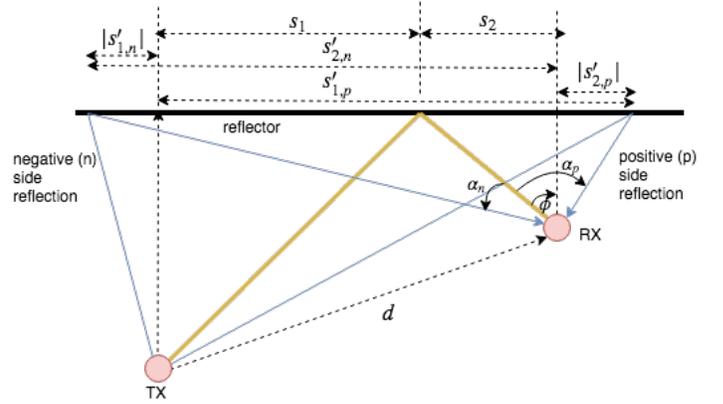} \\
\caption{Diagrams for the cases diffuse ray reflects from out of the normals frame for positive (with $\alpha_p$) and negative (with $\alpha_n$) reflections}\label{outcases}
\end{figure}

In this subsection, we check the other scenarios of which given equations so far were not considering in the analytical setup. These scenarios can be basically defined as the reflections occur outside of the reflector normal frame which is demonstrated in Fig. \ref{deterministic}. %The diagram in Figure \ref{wallnormal} shows the interested area where the reflection occurs at as well as the notations of the terminologies that are used for reflector normals.
In Fig. \ref{basicmodel} and \ref{txbw}, other scenarios are shown as Case 2 for reflector length and transmitter beamwidth calculation, respectively, while Fig. \ref{outcases} is given for path length calculations. We claim that the setup formulations in the previous sections are still valid and refer the reader to Appendix \ref{appen_formulvalid} for the proofs.

 %\begin{figure}[t]
%\centering
%\includegraphics[scale=0.55]{wallnormal.png} \\
%\caption{Diagram of the interested area to check the model setup}\label{wallnormal}
%\end{figure}

	\section{Intra-Cluster Channel Model Setup using Basic Geometric Model} \label{sub_IntraClusterModel}
	
%According to the Basic Geometric Model given in the previous section, given the specified environmental parameters, we are able to determine the time of arrival and angle of arrival of any ray received through a first-order reflection. 
In this section, using the Basic Geometric Model, we generate a first-order reflection cluster structure that consists of multiple rays as defined in Section ‎\ref{sub_modelCluster}. Throughout the paper, we assume that both the channel and the transceiver-receiver pair are stationary which means the channel impulse response is time-invariant. 

To estimate the channel parameters, we propose a \textit{deterministic approach} where we let infinitely many rays depart from the transmitter. %, reflect and reach to the receiver within the support region. We calculate the delay for all rays. In addition to angle and time parameters, amplitude and the phase of each ray is also calculated. The calculations of power and phase are given in Sections ‎\ref{subsub_PowerCalc} and ‎\ref{subsub_PhaseCalc}.
In particular, in this section, we will introduce the deterministic model setup, give a theoretical cluster channel impulse response (TC-CIR), calculate its parameters. Finally, we study how to bin the resultant profiles to get the practical multipath channel impulse response. To do so, we create a novel mmwave spatial channel model.

		\subsection{System Setup}\label{subsub_infiniteRays}
		
Fig. \ref{deterministic} illustrates the infinitely many rays approach. In this approach, the number of rays within the visible region is assumed to be infinity. To approximate the infinity number of rays, we digitize the support range with very small spacing ($\Delta\alpha$). So, the number of rays in digitized spatial domain is $N_r^d= \left\lfloor (\alpha^- - \alpha^+)/\Delta\alpha \right\rfloor$ where $\alpha^+$ and $\alpha^-$ are given in Table \ref{supportrange}. Then, the offset AoA of $k$-th ray in the cluster is $\alpha_k =(\alpha^-)+k\Delta\alpha$ where $k=0, 1, \hdots, N_r^d-1$, excluding the specular ray offset AoA of 0, i.e. $\alpha_{sp}=0$. 

With these definitions, the BGM can be applied directly. The method scans all $\alpha$ values within the support range with $\Delta\alpha$ increments. For every $\alpha_k$, it calculates the $\tau_k$, the delay of $k$-th ray within the cluster. Hence, the length and delay for the $k$-th ray in the cluster can be given as 
\begin{equation}
\begin{split}
l_k=&\frac{h_r}{\cos (\phi-\alpha_k)} \\
&+\sqrt{h_t^2+ \left( s-\frac{h_r}{\cos (\phi-\alpha_k)}\sin (\phi-\alpha_k) \right) ^2}
\end{split} \label{lk}
\end{equation}

and $\tau_k=t_k-t_{sp}$ where $k=0, 1, \hdots, N_r^d-1$, $t_{sp}$ is the ToA of the specular reflection already defined in BGM  and $t_k=l_k/c$ is the ToA of the $k$-th ray.

As a result, the baseband theoretical cluster channel impulse response (TC-CIR) becomes
\begin{equation}
\begin{split}
c_T(t_{sp},\phi)&=a_{sp} e^{j\varphi_{sp}}\delta (t_{sp})\delta (\phi)\\
&+\sum_{k=0}^{N_r^d-1} a_k e^{j\varphi_k} \delta (t_{sp}-\tau_k) \delta (\phi-\alpha_k) \label{CIR_theo}
\end{split}
\end{equation}

where $a_{sp}$ and $\varphi_{sp}$ are the amplitude and the phase of the specular ray; $a_{k}$, $\varphi_{k}$, $\tau_{k}$, $\alpha_{k}$ are amplitude, phase, delay, offset AoA of the $k$-th ray, respectively. $\delta (.)$ is  Dirac delta function and $N_r^d$ is the number of rays. Note that $c_T(t_{sp},\phi)$ is the function of time and angle of arrival of the specular ray. In section ‎\ref{subsub_PowerCalc} and ‎\ref{subsub_PhaseCalc} we will give the formulation for estimating the amplitudes $a_k$ and phases $\varphi_k$ for the $k$-th ray.

		\subsection{Directive Diffuse Scattering Model} \label{subsub_DiffuseScatMod}

In mmWave channels, even very tiny variations in a typical reflector create scattering since the wavelength is very small \cite{11ay, Maltsev_office, MiWEBA}. According to the measurement results at 60 GHz given in \cite{Gentile}, received power due to the diffuse scattering was as high as $26\%$ of the total cluster power. Apparently,
%In Section ‎\ref{Sec_diffuse}, we stated that the 
diffuse scattering is a non-negligible propagation mechanism in mmWave channels and hence has to be taken into account when modeling the cluster channels. %We already introduced the Basic Geometric Model that takes a diffuse ray into account and calculates the angle between the specular ray and the diffuse ray at the receiver end. However, it is also described in Section ‎\ref{Sec_diffuse} that diffuse scattering depends on the surface roughness, incident angle and the carrier frequency. If the surface is highly rough, the scattering is assumed to be towards every direction including the back of the incident signal. This phenomenon is replicated with the \textit{Lambertian Model} \cite{Glassner}, and its roots are at optics as the visible light frequency is on the order of terahertz where almost all objects are rough. However, in mmWave, the scattering direction range is limited by the factors given above. That is, not every diffuse ray may reach to the receiver even if it is still in the support region that is described in Section ‎\ref{subsub_supportReg}. As a result, the 
The angular shape of the scattering event should also be modeled in order to estimate the directions (as well as the relative powers with respect to the specular ray) of the diffuse rays.

In \cite{Esposti}, scattering event is modeled with 3 different patterns. According to the measurements, the directive pattern is the most accurate model and given in our context as
\begin{equation}
\rho_k(\psi_k,m)=\left(\frac{1+\cos \psi_k}{2}\right)^m \label{rho_k}
\end{equation}

where $\rho_k(\psi_k,m)$ is defined to be the relative diffuse scattering power coefficient of the $k$-th diffuse ray with respect to its specular reflected ray. It is a function of $\psi_k$ and $m$ where $\psi_k$ is the angle between the specular reflected ray and the diffuse reflected ray of the $k$-th diffuse ray and $m$ is the design parameter that determines the width of the pattern. Note that the function takes its maximum at $\psi_k=0$, i.e. specularly reflected ray of the $k$-th diffuse ray. And $\rho_k(0,m)=1$ for any $m$. Also, we assumed $m$ to be equal for all $k$; meaning that the roughness of the surface is same everywhere and doesn't depend on the grazing angle. 
%To avoid confusing, we slightly change the notation on the scattering shape given in Eq. (\ref{scatpat}): 
 
%We defined the Directive ER model in Section ‎\ref{sub_ER} where scattering directions are modeled with a directive pattern in Eq. (\ref{scatpat}). l for mmWave frequencies. Hence, we adopt the pattern to model the scattering at the surface and embed it into our geometrical ray tracing algorithm. To do so, the overall intra-cluster model is updated and displayed in the diagram in Figure \ref{diffusemod} for $N_r^d=6$.
 \begin{figure}[t]
\centering
\includegraphics[scale=0.5]{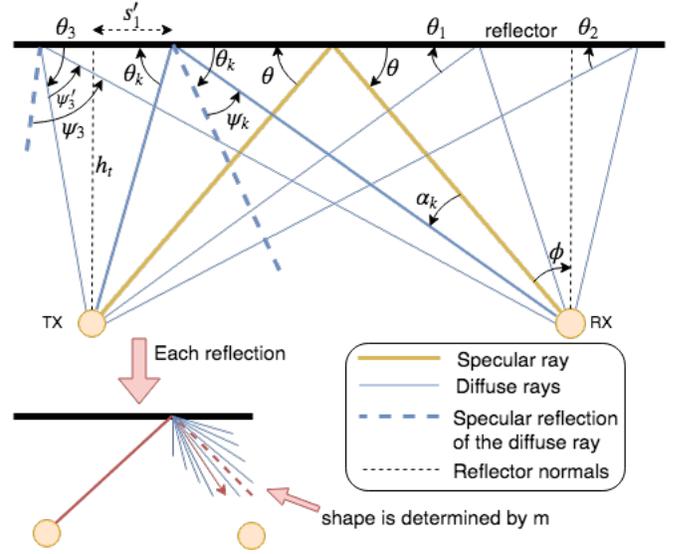} \\
\caption{Updated cluster model diagram with the addition of diffuse scattering pattern}\label{diffusemod}
\end{figure}

As the Fig. \ref{diffusemod} demonstrates, scattering is assumed to occur at each reflection point of the incident ray, and only one ray in the scattered pattern can reach to the receiver. Also note that each incident ray has its own grazing angle, $\theta_k$. In this context, BGM needs to be updated too. Consider the diffuse ray with the grazing angle $\theta_k$ in Fig. \ref{diffusemod}. Since only one reflected ray reaches the receiver, we only take one direction into account within the scattering pattern. In order to calculate the offset direction ($\psi_k$) of the $k$-th diffuse ray, first we need to find the grazing angle associated that ray, i.e. $\theta_k$. From Snell's Law, grazing angle of the incident ray equals to the grazing angle of the reflected specular.% ray which is shown in Figure \ref{diffuseupdate} too.
% \begin{figure}[t]
%\centering
%\includegraphics[scale=0.55]{diffuseupdates.png} \\
%\caption{Diagram of diffuse scattering parameters addition on Basic Geometric Model}\label{diffuseupdate}
%\end{figure}
Then, % the grazing angle of the diffuse ray is
\begin{equation}
\theta_k=\tan^{-1}\frac{h_t}{|s_1^{\prime}|} \label{thetak}
\end{equation}

where $s_1^{\prime}$ is given in Appendix \ref{appen_calculate}. $\psi_k$ and $\alpha_k$ are negative in the diagram. Hence, 
\begin{equation}
\psi_k=90-(\phi-\alpha_k)-\theta_k \label{psik}
\end{equation}

We claim that the formulas are valid for all cases. The proof is in Appendix \ref{appen_diffuseValidation}.
 
		\subsection{Power Calculation of the Rays} \label{subsub_PowerCalc}

%In wireless communications, there are several propagation mechanisms apply to the transmitted signal over the space that degrade the signal power. In this section, we will extract the power spectrum of the proposed channel model by taking effective loss mechanisms into account in mmwave frequencies. We already discussed that, in mmWave channels, the most important propagation loss is due to the free space attenuation so we will start with formulation of free space path loss. On the other hand, since the model is setup based on the first-order reflection, reflection loss will be analyzed. Furthermore, we stated that the scattering has a non-negligible effect on mmwave channels which suggests an increased loss at the reflection point due to scattering that has to be accounted as well. Using both the ITU Model and Directive Model, we study the scattering loss. Finally, for 60 GHz, we calculate the oxygen absorption loss too.

%As described in the beginning of Section \ref{sub_IntraClusterModel}, each ray in the theoretical impulse response will encounter different amplitude and phase values as the propagation mechanisms apply to each ray, individually. For that reason, the approach is to calculate the power of each ray separately, based on the function of the propagation mechanism that will be generalized in the corresponding sections.

We define the transmission equation for the $k$-th ray such that the received ray power is given as
\begin{equation}
P_k=\frac{P_T G_T G_R}{L_k R_k S_k} \label{Pk}
\end{equation}

where $P_T$ is the transmit power; $G_T$ and $G_R$ are the transmitter and receiver antenna gain, respectively; $L_k$, $R_k$ and $S_k$ are the losses applied to the $k$-th ray due to free space, reflection and scattering, respectively. In the following sections, we give an expression for $L_k, R_k, S_k$.

\subsubsection{Free Space Loss} \label{SSS_FSPL}

%Although free-space path loss is formulated for modeling the loss due to the free space that the signal travels within, we will apply the same model to our case where signal reflects from a reflector which violates the free space requirement. Since we accommodate the reflection effects within the Reflection Loss section, this assumption is safe. Then 
The path loss applied to the $k$-th ray in linear scale is 
\begin{align}
L_k&=\left( \frac{\lambda}{4\pi l_k}\right)^2, & k=0, 1, \hdots, N_r^d-1
\end{align}

where $\lambda$ is wavelength of the carrier frequency and $l_k$ is the length of the ray that is given in Eq. (\ref{lk}). %We proved that the shortest ray in the cluster is the specular ray in Remark 2. Since shorter length implies the earlier arrival, as seen in the theoretical channel impulse response in Eq. (\ref{CIR_theo}), $\tau_{sp}=0$ and $\alpha_{sp}=0$. Then, using Remark 2, following inequality can be given:
%\begin{equation}
%L_{sp}<L_k
%\end{equation}

%for any $ k=0, 1, \hdots, N_r^d-1$. 

\subsubsection{Reflection Loss} \label{SSS_RL}

%When an EM wave reflects from a dielectric material, due to the imperfect conductivity of the material, incident energy is reflected back with a loss, which we will call as \textit{reflection loss}. 
Reflection loss applied to the incident electric field can be characterized through the \textit{Fresnel reflection coefficient} ($\Gamma$)  \cite{Orfanidis}. %that depends on the material properties, electric field polarization and the grazing angle and can be related to the incident power as
%\begin{equation}
%|\Gamma|^2=\frac{P_r}{P_i}
%\end{equation}
%where $P_i$ and $P_r$ are the incident and reflected signal powers, respectively.
There are two Fresnel equations for two polarization cases to calculate the Fresnel coefficient ($\Gamma$). The simplified versions of the equations for vertically and horizontally polarized $k$-th ray in our model are given as, respectively,
\begin{equation}
\Gamma_k^{\parallel}=\frac{-\varepsilon_r \sin \theta_k + \sqrt{\varepsilon_r-\cos^2(\theta_k)}}{\varepsilon_r \sin \theta_k + \sqrt{\varepsilon_r-\cos^2(\theta_k)} } \label{Rvert_coef}
\end{equation}
\begin{equation}
\Gamma_k^{\perp}=\frac{\sin \theta_k - \sqrt{\varepsilon_r-\cos^2(\theta_k)}}{\sin \theta_k + \sqrt{\varepsilon_r-\cos^2(\theta_k)}} \label{Rhor_coef}
\end{equation}

where $\theta_k$ is the grazing angle defined in Eq. (\ref{thetak}) and $\varepsilon_r$ is the relative permittivity of the reflection material that is a given parameter through the measurements. % and its value for several different materials are given in Table 3. 
It is also worthy to note that $\varepsilon_r$ doesn't depend on the carrier frequency \cite{Rappaport_book, ITU2}.
%\begin{table}[!t]
%\centering
%\caption{Relative permittivity of some materials \cite{ITU2}} 
%\label{relativeperm}
%\begin{tabular}{|c|c||c|c|}
%\hline
%Material  &  $\varepsilon_r$ & Material &  $\varepsilon_r$
 % \\ \hline \hline
%Concrete & 5.31 & Metal & 1  
%\\ \hline
%Brick & 3.75 & Very dry ground & 3
%\\ \hline
%Plasterboard & 2.94 & Medium dry ground & 15 
% \\ \hline 
%Wood & 1.99 & Wet ground & 30                                                                                                           
%\\ \hline
%Glass  & 6.27 & Chipboard & 2.58                                                                                                           
%\\ \hline
%Ceiling board & 1.5 & Floorboard & 3.66                                                                                                           
%\\ \hline
%\end{tabular}
%\end{table}

As a result, reflection loss coefficients in linear scale for $k$-th ray are defined as
%\begin{equation}
%R_k^{\parallel}=\frac{1}{|\Gamma_k^{\parallel}|^2}=\left|\frac{\varepsilon_r \sin \theta_k + \sqrt{\varepsilon_r-\cos^2(\theta_k)}}{-\varepsilon_r \sin \theta_k + \sqrt{\varepsilon_r-\cos^2(\theta_k)}} \right|^2 \label{Rkparallel}
%\end{equation}
%\begin{equation}
%R_k^{\perp}=\frac{1}{|\Gamma_k^{\perp}|^2}=\left| \frac{\sin \theta_k + \sqrt{\varepsilon_r-\cos^2(\theta_k)}}{\sin \theta_k - \sqrt{\varepsilon_r-\cos^2(\theta_k)}} \right|^2 \label{Rkperp}
%\end{equation}

%Then, 
\begin{align}
R_k=
\left\{
	\begin{array}{ll} \vspace{1.5mm}
		R_k^{\parallel}=1/|\Gamma_k^{\parallel}|^2  & \mbox{if vertical pol.}\\
		R_k^{\perp}=1/|\Gamma_k^{\perp}|^2 & \mbox{if horizontal pol.}
	\end{array}
\right.
\end{align}

%For horizontal polarization, Eq. (\ref{Rkperp}) outputs an increasing loss while grazing angle increases. Hence, as seen from Figure \ref{diffusemod}, since some rays have smaller grazing angle compared to that of specular ray, we may conclude that some of the diffuse rays encounter a reflection loss lower than specular ray reflection loss.

%For vertical polarization, on the other hand, while some diffuse rays may still have lower reflection loss, Eq. (\ref{Rkparallel}) is not monotone increasing as angle increases. The loss coefficient ($|\Gamma_k^{\parallel}|^2$) reaches 0 (infinity loss) at an angle regardless of $\varepsilon_r$ and then increases back for further increasing grazing angle. At that angle, called \textit{Brewster Angle}, whatever power incident ray has, nothing is reflected back into the free space. 

\subsubsection{Scattering Loss} \label{SSS_SL}

The scattering loss is studied in \cite{ITU} and the loss coefficient for the specular component is given as
\begin{equation}
\rho_s(\theta)=\exp \left( -0.5 \left( \frac{4 \pi \sigma_h}{\lambda} \sin \theta \right)^2 \right) \label{rho_s}
\end{equation}

where %$g$ is given in Eq. (\ref{g}). %However, it is claimed in the report that there is no exact expression for the diffuse component and hence its loss coefficient, $\rho_d$, is modeled as a random variable which is not suitable at our deterministic model.  
$\sigma_h$ is the standard deviation of the surface height ($h$) about the local mean within the first Fresnel zone, $\lambda$ is the carrier wavelength and $\theta$ is the grazing angle. Here, variations on the surface, or surface height, $h$, is modeled as a Gaussian distributed random variable \cite{Rappaport_book}.

On the other hand, for the $k$-th incident ray, a relation between the power degradation at $k$-th specular ray and its any diffuse ray is given in Eq. (\ref{rho_k}) via a scattering pattern.

Hence, the scattering loss for the $k$-th ray can be given as
\begin{equation}
S_k=\left( \frac{1}{\rho_{s,k}\rho_k}\right)^2 \label{Sk}
\end{equation}

where $\rho_{s,k}$ is the specular ray coefficient that expresses the loss applied to the $k$-th incident ray caused by the roughness of the material and is given by $\rho_{s,k}=\rho_s(\theta_k)$ and $\rho_k$ is the coefficient that accounts the loss due to the power dispersion after scattering given in Eq. (\ref{rho_k}). 

%\paragraph{Oxygen Absorption Loss} \label{SSS_OxygenL}

%At 60 GHz carrier frequency, additional loss is applied to the radio wave due to the oxygen absorption \cite{MiWEBA} %It is already expressed with a function that depends only on the distance that the wave travels and a fixed multiplying constant . Absolute oxygen absorption loss at 60 GHz frequency
%and is given for the $k$-th ray as 
%\begin{equation}
%A_k=10^{0.0015}l_k
%\end{equation}

%As a result, Eq. (\ref{Pk}) is updated for 60 GHz as
%\begin{equation}
%P_k=\frac{P_T G_T G_R}{L_k R_k S_k A_k} \label{Pk_60G}
%\end{equation}

%In dB scale,
%\begin{equation}
%\begin{split}
%P_k^{dB}=&P_T^{dB}+ G_T^{dB}+ G_R^{dB} \\
%&- L_k^{dB} - R_k^{dB} - S_k^{dB} - A_k^{dB} \label{Pk_dB}
%\end{split}
%\end{equation}

%where $X^{dB}=10\log X$. For carrier frequencies other than 60 GHz, $A_k^{dB}=0$.

Finally, since we introduce amplitudes in the cluster channel model in Eq. (\ref{CIR_theo}), power can be converted to absolute value of amplitudes via $|a_k|=\sqrt{P_k}$.

		\subsection{Phase Calculation of the Rays} \label{subsub_PhaseCalc}
		
Rays arrive receiver with different phases due to the difference at their path length and at the grazing angle during the reflection. Hence, in order to be able to sum the ray powers properly, phase information of each ray should be calculated deterministically.

\subsubsection{Phase Offset due to Path Distances} \label{SSS_PhaseD}

The phase offset of the $k$-th ray due to the path difference with respect to the specular ray can be given as $\Delta \varphi_{D,k}=(2 \pi (l_k-l_{sp}))/\lambda$.

%Note that from Eq. (\ref{phased}), a path length difference of half wavelength, i.e.$l_k-l_{k-y}=\lambda/2$ , results in an out-of-phase difference, where two rays add each other, destructively. Considering the mmWave wavelength of $5$ mm for 60 GHz, that corresponds to $2.5$ mm path difference where the destructive summation phenomenon might occur easily. That shows how sensitive the mmWave channels to the phase offsets caused by path length difference.

\subsubsection{Phase Offset due to Reflection} \label{SSS_PhaseR}

%The other propagation mechanism that changes the phase of the rays is reflection. 
Note that Fresnel equations given in (\ref{Rvert_coef}) and (\ref{Rhor_coef}) are complex coefficients. %Along with the amplitude change, they reveal the phase changes applied to the electric field at the reflection point as well. 
Hence, we can define the phase offset introduced by the reflection to the $k$-th ray as
\begin{align}
\Delta\varphi_{R,k}^{\prime}=
\left\{
	\begin{array}{ll} \vspace{1.5mm}
		\angle  \Gamma^{\parallel}  & \mbox{if vertical pol.}\\
		\angle  \Gamma^{\perp} & \mbox{if horizontal pol.}
	\end{array}
\right. \label{phaser}
\end{align}

In order to be able to calculate a total instant phase of a ray, we need to align with the same reference (specular ray phase offset) with the previous subsection. Hence, we correct the phase offset of the $k$-th ray due to the reflection with respect to the specular ray as $\Delta\varphi_{R,k}=\Delta\varphi_{R,k}^{\prime}-\Delta\varphi_{R,spec}^{\prime}$ %\label{phaseRk}
%\end{equation}

%\paragraph{Instant Ray Phase at the Receiver} \label{SSS_PhaseInstant}
Overall, phase offset of the $k$-th ray with respect to the specular ray can be given as 
\begin{equation}
\Delta\varphi_k=\Delta\varphi_{D,k}+\Delta\varphi_{R,k}
\end{equation}

%where $\Delta\varphi_{D,k}$ is the path difference contribution given in Eq. (\ref{phasedk}) and $\Delta\varphi_{R,k}$ is the reflection contribution given in Eq. (\ref{phaseRk}). 
Note that $\Delta\varphi_k=\varphi_k-\varphi_{sp}$. Further, assuming $\varphi_{sp}=0$, $\Delta\varphi_k=\varphi_k$.

		\subsection{Binned Intra-Cluster Channel Model} \label{subsub_binning}
		
Since all the rays are not resolvable by the receiver due to the limitation on the resolution, an additional discrete binning is needed on top of the theoretical approach given in Section ‎\ref{subsub_infiniteRays}. 

After binning (filtering and sampling) on both angle and time domains, the resultant discrete baseband time-invariant channel impulse response for the cluster (C-CIR) can be given as following:
\begin{equation}
c[n_{sp}, \Omega_{sp}]=\sum_{i=0}^{N_r-1} a^{(i)} e^{j\varphi^{(i)}} \delta [n_{sp}-i\Delta\tau] \delta [\Omega_{sp}-i\Delta\phi] \label{CIR}
\end{equation}

where $n_{sp}$ and $\Omega_{sp}$ are the ToA and AoA of the specular ray in discrete time and angle domain. $\Delta\tau$ and $\Delta\phi$ are the time and angle resolutions; $a^{(i)}$ and $\varphi^{(i)}$  are amplitude and phase of the $i$-th MPC, respectively and $N_r$ is the number of multipath components (MPCs). Finally, $\delta[.]$ is the Kronecker delta function.

Eq. (\ref{CIR}) defines the space-time characteristics of a cluster that consists of $N_r$ MPCs. In other words, it maps the components from the time domain to the angle domain (and vice versa) with respect to specular ray. $N_r$ is a system-dependent parameter that depends on the receiver, the signal bandwidth and/or the type of the measurement. %For example, if the receiver has an omnidirectional antenna, then spatial information of the channel cannot be collected; thereby the channel impulse response given in Eq. (\ref{CIR}) reduces to the temporal channel model only. On the other hand, if a directional antenna is used at the receiver, the channel model of the \textit{beamformed} link also doesn't include the spatial information in MPC level. An example to that case is the intra-cluster model given in \cite{Maltsev_office}. Experiment is held such a way that both the transmitter and the receiver beams are steered towards the specular ray direction of the first-order reflection cluster. 7 to 10 rays are captured in time domain, i.e. $N_r=7$-$10$.
%However, 
Using either the \textit{channel sounding with directional measurements} technique or beamforming, spatial information of the overall channel can be extracted. In this way, timing characteristics for each angle resolution can be assigned. So, the 2-D (spatial-temporal) channel model given in Eq. (\ref{CIR}) is validated. 

In the binning procedure from Eq. (\ref{CIR_theo}) to (\ref{CIR}), we assume that the angular resolution is determined by scanning increments ($\Delta\phi$) in measurement technique and by receiver beamwidth ($\Theta_r$) in beamforming. We assume the channel is narrowband for each angle resolution where the bandwidth of the channel is larger than the signal bandwidth. Then there is single MPC in time domain for each angle resolution. Mapping in both domains is performed such a way that the MPC is located in the middle of the bin while the MPC power is obtained via phasor summation of the ray powers within the bin. Finally, MPCs that have power lower than the receiver sensitivity should be discarded. That is, whenever  $10 \log |a^{(i)}|^2< P_{RS}$ where $P_{RS}$ is the receiver sensitivity, the MPC is removed from the C-CIR. 

\begin{figure}[t]
\centering
\includegraphics[scale=0.45]{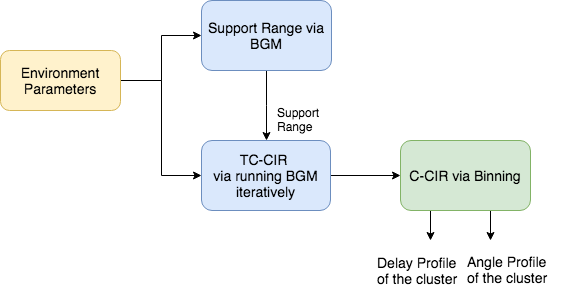} \\
\caption{Flowchart diagram of the C-CIR generation.}\label{flowchart}
\end{figure}

To summarize the overall proposed model in the paper, a flowchart given in Fig. \ref{flowchart} shows the operations to obtain the time and angle domain representations of the cluster for a specified communication system. 

%		\subsubsection{Total Cluster Power} \label{Sec_TotalClusterPower}
		
%From \cite{Rappaport_book}, total cluster power is simply given as 
%\begin{equation}
%P_{cl}=\sum_{i=0}^{N_r-1} |a^{(i)}|^2
%\end{equation}
		\section{Extension to MIMO} \label{Sec_MIMO}	
		
		\subsection{Channel Impulse Response}
		
In this paper, we created a channel model for the cluster in the channel. However, the extension model that covers the overall channel can also be introduced. As a generic model, if receiver antenna has an omnidirectional antenna pattern, the discrete channel impulse response that only considers single-order reflection clusters is given as
\begin{equation}
h[n,\Omega]=\sum_{j=0}^{N_{cl}-1} c^{(j)}[n-T^{(j)}, \Omega-\Phi^{(j)}] \label{CIR_channel}
\end{equation}

where $n, \Omega$ is time and angle of arrivals at the receiver; $T^{(j)}$ and $\Phi^{(j)}$ are delay and AoA of the $j$-th cluster; $N_{cl}$ is the number of clusters and $c^{(j)}$ is discrete channel impulse response of the $j$-th cluster given in Eq. (\ref{CIR}).

Note that if we define the \textit{specular ray} ToA and AoA of a cluster to be the \textit{cluster} ToA and AoA, assigning $n=0$, $T^{(j)}=n_{sp}^{(j)}$ where $n_{sp}^{(j)}$ is the ToA of the $j$-th cluster. In order to create a similar relation in angle domain, $\Omega$ should be fixed for all clusters. However, we setup the BGM model assuming the reference direction is the reflector normal at the receiver (RNR) and $\phi$ is called specular ray AoA with respect to RNR. Apparently, reference direction changes for each cluster. Here, we define LOS ray AoA as the new reference which would be fixed for any first-order reflection scenarios within the channel. The transformation from $\phi$ to $\Phi$ is given as $\Phi=90-\phi + \sigma$ where $\Phi$ is the AoA of the specular ray within the cluster with respect to LOS ray; $\sigma$ is introduced in subsection \ref{SSS_limitGeo} and its visualization can be seen in Fig. \ref{basicmodel}. Hence for $\Omega=90+\sigma^{(j)}$ where $\sigma^{(j)}$ is the tilt angle of $j$-th cluster, $\Phi^{(j)}=\phi^{(j)}$, i.e. $\phi^{(j)}$ is the AoA of the $j$-th cluster.

\subsection{SISO Channel Impulse Response}

So far, we did not state an explicit constraint whether the link between the transmitter and the receiver is beamformed. However, when support region for $\alpha$ was being calculated in BGM setup, transmit beamforming is implicitly accounted by taking the transmitter beamwidth into account. Also, the aim of the overall paper was stated as to give an angle domain presentation of the cluster in order to minimize misalignments of the receiver beams. Then, for a single-input-single-output (SISO) NLOS scenario, if the receiver antenna is beamformed to the provided cluster direction, then the obtained cluster channel impulse response (C-CIR) given in Eq.(\ref{CIR}) becomes the channel impulse response (CIR) and given as
\begin{equation}
h[n,\Omega]=c[n-T,\Omega-\Phi] \label{SISO}
\end{equation}

\subsection{LOS Ray Parameters}

Note that if the scenario is LOS, we will consider that one of the clusters will be through LOS, i.e. no reflection. As in other ray tracing algorithms, we model that cluster with the single LOS ray. In this case, for $n=0$ in Eq. (\ref{CIR_channel}), $T=n_{los}$, i.e. LOS ToA in discrete time domain. For $\Omega=0$, LOS ray AoA $\Phi=0$. Finally, its power in linear scale is given by $P_{los}=(P_T G_T G_R/L_{los})$ 
where $L_{los}$ is the LOS ray attenuation due to the free space path-loss and given as $L_{los}=\left( \lambda/4\pi d \right)^2$. 

\subsection{MIMO Channel Impulse Response}
		
\begin{figure}[t]
\centering
\includegraphics[scale=0.35]{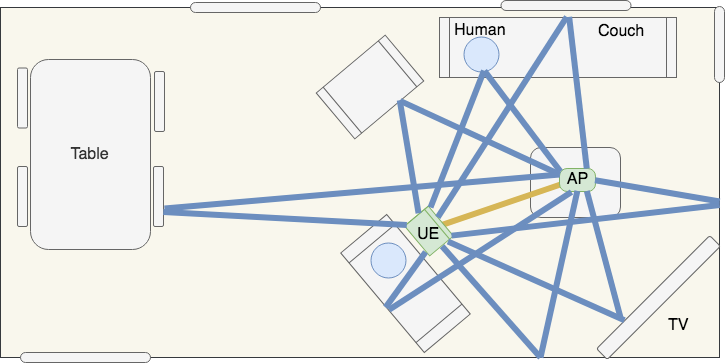} \\
\caption{Possible single-order clusters in a MIMO communication within a typical living room environment}\label{MIMO}
\end{figure}

In BGM setup, we discuss the cluster angle spread limitation due to transmit beamwidth. In mmWave MIMO,  due to the large array usage opportunity, antenna beamwidth can reach to very small values (smaller than $1^{\circ}$ with $64$ antenna elements \cite{Orfanidis}) which makes the transmit beamwidth dominant limitation factor in our intra-cluster model, especially for indoor environments. Same phenomenon occurs in outdoor mmWave applications if massive MIMO is used in the communication system where even smaller beamwidth can be achieved. This is because $h_t$ is also a factor that determines the beamwidth limitation and it should be relatively small to keep the transmit beamwidth as a dominant limiting factor. As a result, several spatially separated single-order reflection clusters are provided within the channel via beamformed links.  Fig. \ref{MIMO} gives an example for a typical living room. 

Proceeding with the consideration of a dedicated transmit-receive beamformed link for each cluster, we can think of, in fact, that each cluster constitutes a SISO channel. Then, for $N_{cl}$ beamformed links (clusters), MIMO channel matrix $\mathbf{H}$ can be, analytically, represented as
\begin{equation}
\mathbf{H}=
\begin{bmatrix}
h_{11} & h_{12} & \hdots & h_{1N_{cl}} \\
h_{21} & h_{22} & \hdots & h_{2N_{cl}} \\
\vdots & \vdots & \ddots & \vdots \\
h_{N_{cl}1} & h_{N_{cl}2} & \hdots & h_{N_{cl}N_{cl}} \\
\end{bmatrix}
\end{equation}
where $h_{pq}$ is the channel impulse response for the link between $p$-th transmit beam and $q$-th receive beam. Apparently, intended beamformed links are denoted for $p=q$ cases whereas the interference between the links are shown as $p \neq q$.
 
In this paper, we consider the clusters are perfectly separated in the spatial domain. That is, the beams aligned to the different first-order reflection directions do not overlap each other. With this assumption, the extension to MIMO becomes straightforward and the channel matrix reduces to
\begin{equation}
\mathbf{H}=
\begin{bmatrix}
h_{11} & 0 & \hdots & 0 \\
0 & h_{22} & \hdots & 0 \\
\vdots & \vdots & \ddots & \vdots \\
0 & 0 & \hdots & h_{N_{cl}N_{cl}} \\
\end{bmatrix}
\end{equation}

where $h_{jj}$ is given in Eq. (\ref{SISO}) for $j=1,2, \hdots, N_{cl}$.

\subsection{Massive MIMO and Intra-Cluster Model}

Now that the proposed model provides the detailed spatial representation of the MIMO channel in the cluster level, several novel beamforming techniques can be introduced using massive MIMO approaches to increase the spatial usage of the channel. In this section, we give the insights of three, but several others can be introduced that exploit the proposed model. 

\paragraph{Adjusting the Optimum Beamwidth}

From array processing techniques, we already know that increased beam gain can be achieved by narrowing the beam. Since the number of antennas used to create the beam also determines the beam gain and beamwidth \cite{Orfanidis}, now one can calculate the optimum beamwidth and select the number of antennas within the port and create the desired beamwidth. 

\paragraph{Different Beamwidth for Each Cluster}

As we will show in the implementation section, clusters have different angle spreads which implies that each beam dedicated to a cluster has its own beamwidth. Considering the large number of antenna elements availability in massive MIMO systems, by adaptively selecting the number of antennas within the antenna ports, beams with different beamwidths can be organized easily. 
\paragraph{Two Beams for Each Cluster} 
Furthermore, the proposed model showed that the angle spectrum of the clusters are not necessarily symmetric. Even for the vertical polarization of the antennas, in some cases, two clusters may be visible from the same first-order reflection. This gives an idea of creating more than one beam with different beamwidths for the same single-order cluster. In that case, another idea could be aligning the beamwidth, not to the specular ray direction but such a way that the beam covers the maximum energy. 

Overall, with the combination of the extensive array processing opportunities of the massive MIMO, the proposed model helps to acquire maximum energy from the channel while increasing the spatial efficiency.
	
\section{Implementation} \label{Sec_implementation}
In this section, we implement the proposed ray tracing channel model, RT-ICM, using the experimental platform performed in \cite{Xu} %and \cite{Maltsev_office} 
and compare the proposed model results with the measurement results. 
For simulation purposes, we set the $N_r^d=1000$ as the larger values don't make any significant difference. %Even $N_r^d=100$ gives accurate results for most of the indoor applications. This is because the receiver can detect approximately 7-10 multipath components at most from a first-order wideband cluster in a typical indoor environment and $N_r^d=100 \gg N_r=10$  in that sense\cite{Maltsev_office}.

\subsection{Indoor 60 GHz - Classroom Environment \cite{Xu}}

 \begin{figure}[t]
\centering
\includegraphics[scale=0.36]{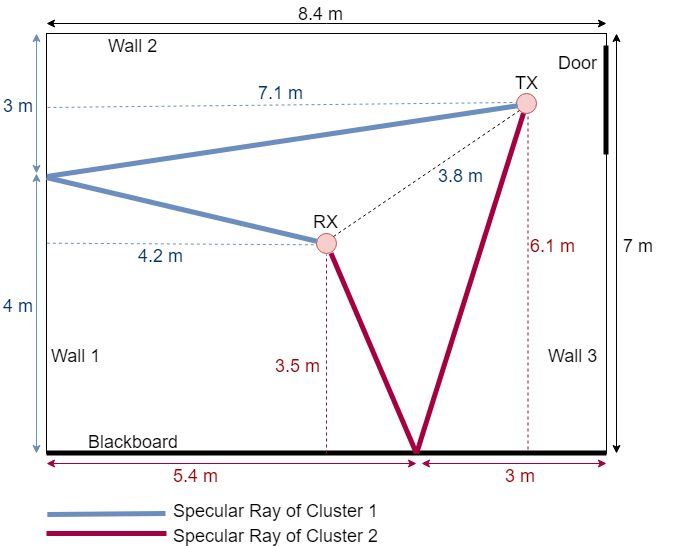} \\
\caption{Measurement environment and the distance parameters when receiver is in the room center.}\label{center}
\end{figure}
In \cite{Xu}, several measurements are held to characterize the spatial and temporal behaviors of the indoor channels at 60 GHz. Totally, 8 experiments are performed in 4 types of environments. Those environments were room, hallway, room to room and corridor to room. Both measurement and statistical results are provided for each measurement. Since the data is collected using spin measurements, power angle profile of the channel is also created. In this paper, we will replicate the measurement environment and the system parameters for the two classroom measurements in \cite{Xu} and implement the proposed channel model.

\subsubsection{Receiver is at the Center}
The top view of the measurement environment is drawn in Fig. \ref{center}. In this first scenario, transmitter is located at the corner and the receiver is in the center of an $8.4 \times 7$ m empty room. One side of the room is covered with a blackboard while the others are indoor building walls. Transmitter has a horn antenna with a beamwidth of $90^{\circ}$ which is directed towards the receiver. And the angle resolution during the spin measurements at the receiver is given as $5^{\circ}$. As a result, although 4 potential reflectors are present within the room, the only possible first-order reflections are through the "Wall-1" and "Blackboard" that are shown in Fig. \ref{center}. Hence, three clusters are considered: cluster 1 and 2 are created through the first-order reflections from wall-1 and blackboard, separately, cluster 3 is the LOS ray. The transmit beamwidth for each cluster is assumed to be $\Theta=45^{\circ}$. We set the diffuse scattering pattern order $m=17$ for the plasterboard wall-1 and $m=35$ for the blackboard (made of slate-stone). Finally, from the measurement result, we choose the power threshold as $P_{RS}=-60$ dBm. The parameters for the clusters are given in Table \ref{inputparams}. 

The resultant theoretical angle domain response of the model is given in  Fig. \ref{centertheoangle}. The ray that arrives with zero AoA is the LOS ray. Resultant specular ray AoA cluster-1 ($\phi^{(1)}$) and cluster-2 ($\phi^{(2)}$) are, respectively, $-127^{\circ}$ and $117^{\circ}$. Angle spread of the clusters are $50^{\circ}$ for cluster-1 and $43^{\circ}$ for cluster-2. The angular spectrum and the numerical values are in agreement with the measurement result figure provided in \cite{Xu} which shows the AoAs as $-128^{\circ}$ and $118^{\circ}$ and the approximate angle spreads as $55^{\circ}$ and $40^{\circ}$.
\begin{figure}[t]
\centering
\includegraphics[scale=0.11]{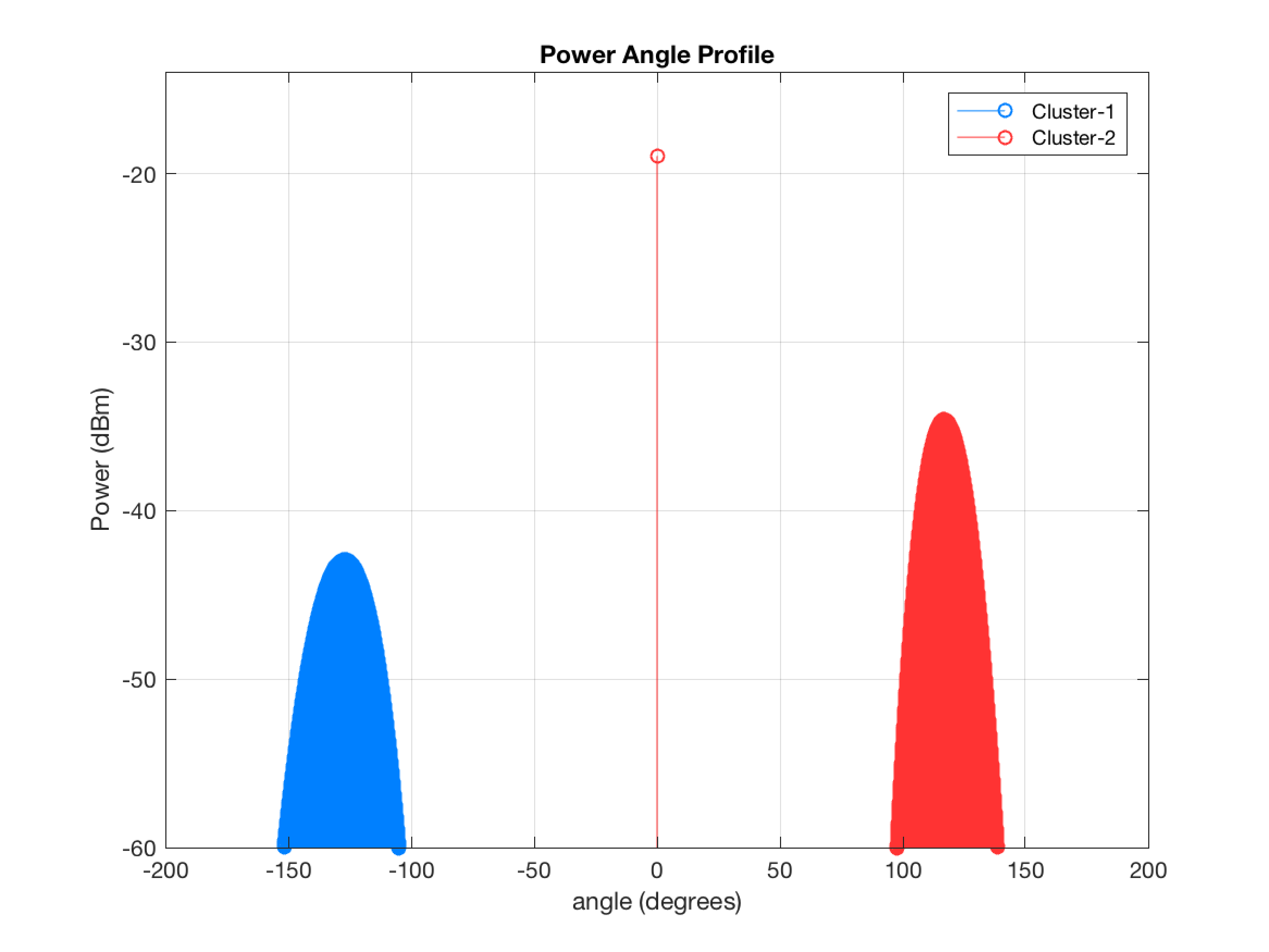} \\
\caption{Theoretical Power Angle Profile of the channel when receiver is in the center of the room.}%\label{centertheoangle}
\end{figure}

\begin{figure}[t]
\centering
\includegraphics[scale=0.55]{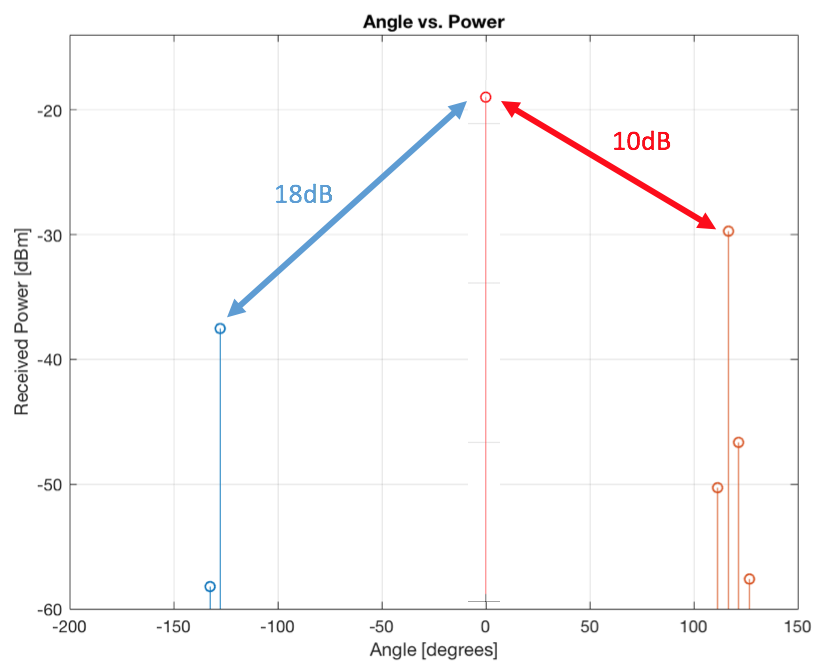} \\
\caption{Binned Power Angle Profile of the channel when receiver is in the center of the room.}\label{centertheoanglebin}
\end{figure}

\begin{table*}[!t]
\centering
\caption{Input Parameters of the Model} 
\label{inputparams}
\begin{tabular}{c|c||c|c|c|c|c|c|c||c|c|c|c}
\hline
Measurements & Clusters & d  & $h_t$ & $h_r$ & $l_{neg}$ & $l_{pos}$ & $\varepsilon_r$ & $\sigma_h$ [mm] &  $P_T$ [dBm] &  $G_T$ [dB] &  $G_R$ [dB] &  $polarization$
  \\ \hline \hline
Room-center & Cluster-1 & 3.8 & 7.1 & 4.2 & 4 & 3 & 2.9 & 0.3 & 25 & 6.7 & 29 & horizontal   
\\ \hline
Room-center & Cluster-2 & 3.8 & 6.1 & 3.5 & 3 & 5.4 & 7.5 & 0.1 & 25 & 6.7 & 29 & horizontal
\\ \hline 
Room-corner & Cluster-1 & 7.1 & 7.1 & 1.2 & 2.2 & 4.8 & 2.9 & 0.3 & 25 & 6.7 & 29 & horizontal
\\ \hline
Room-corner & Cluster-2 & 7.1 & 6.1 & 1.8 & 6.2 & 2.2 & 7.5 & 0.1 & 25 & 6.7 & 29 & horizontal
\\ \hline
%Conference Room & Cluster-1 & 2.6 & 2.1 & 2.1 & 2.25 & 2.25 & 2.9 & 0.2 & 2 & 18 & 18 & horizontal 
%\\ \hline
%Conference Room & Cluster-2 & 2.6 & 0.9 & 0.9 & 2.25 & 2.25 & 2.9 & 0.2 & 2 & 18 & 18 & horizontal 
%\\ \hline
%Conference Room & Cluster-1 & 2.6 & 2.1 & 2.1 & 2.25 & 2.25 & 2.9 & 0.2 & 2 & 18 & 18 & vertical 
%\\ \hline
%Conference Room & Cluster-2 & 2.6 & 0.9 & 0.9 & 2.25 & 2.25 & 2.9 & 0.2 & 2 & 18 & 18 & vertical 
%\\ \hline
\end{tabular}
\end{table*}

On the other hand, to compare the power spectrum in angle domain, power angle profile result after the binning (with $\Delta\phi=5^{\circ}$) is provided in Fig. \ref{centertheoanglebin}. The received powers of cluster-1 and cluster-2 relative to that of LOS ray are $18$ dB and $10$ dB. These also match with measurement results where they were approximately $18$ dB for cluster-1 and $8$ dB for cluster-2.  

\subsubsection{Receiver is on the Corner}
\begin{figure}[t]
\centering
\includegraphics[scale=0.36]{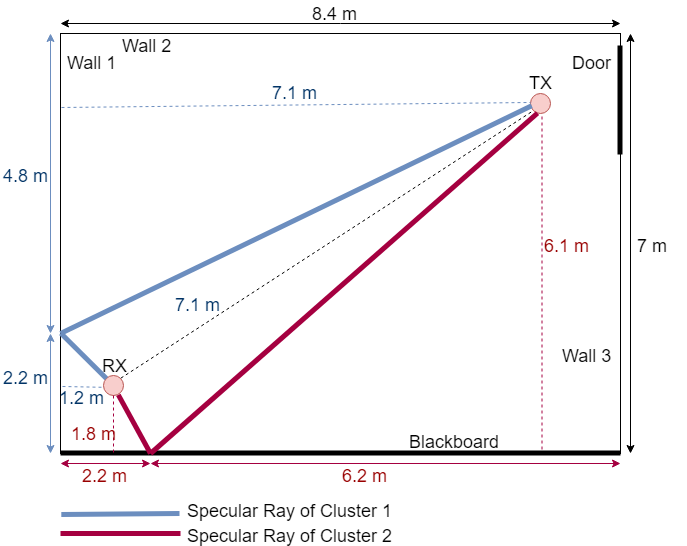} \\
\caption{Measurement environment and the distance parameters when receiver in the room corner.}\label{corner}
\end{figure}

The measurement environment is drawn in Fig. \ref{corner} for the second scenario where the receiver is placed on the corner. The parameters are given in Table \ref{inputparams}. All other parameters that are not listed are the same as in the previous scenario. 

\begin{figure}[t]
\centering
\includegraphics[scale=0.11]{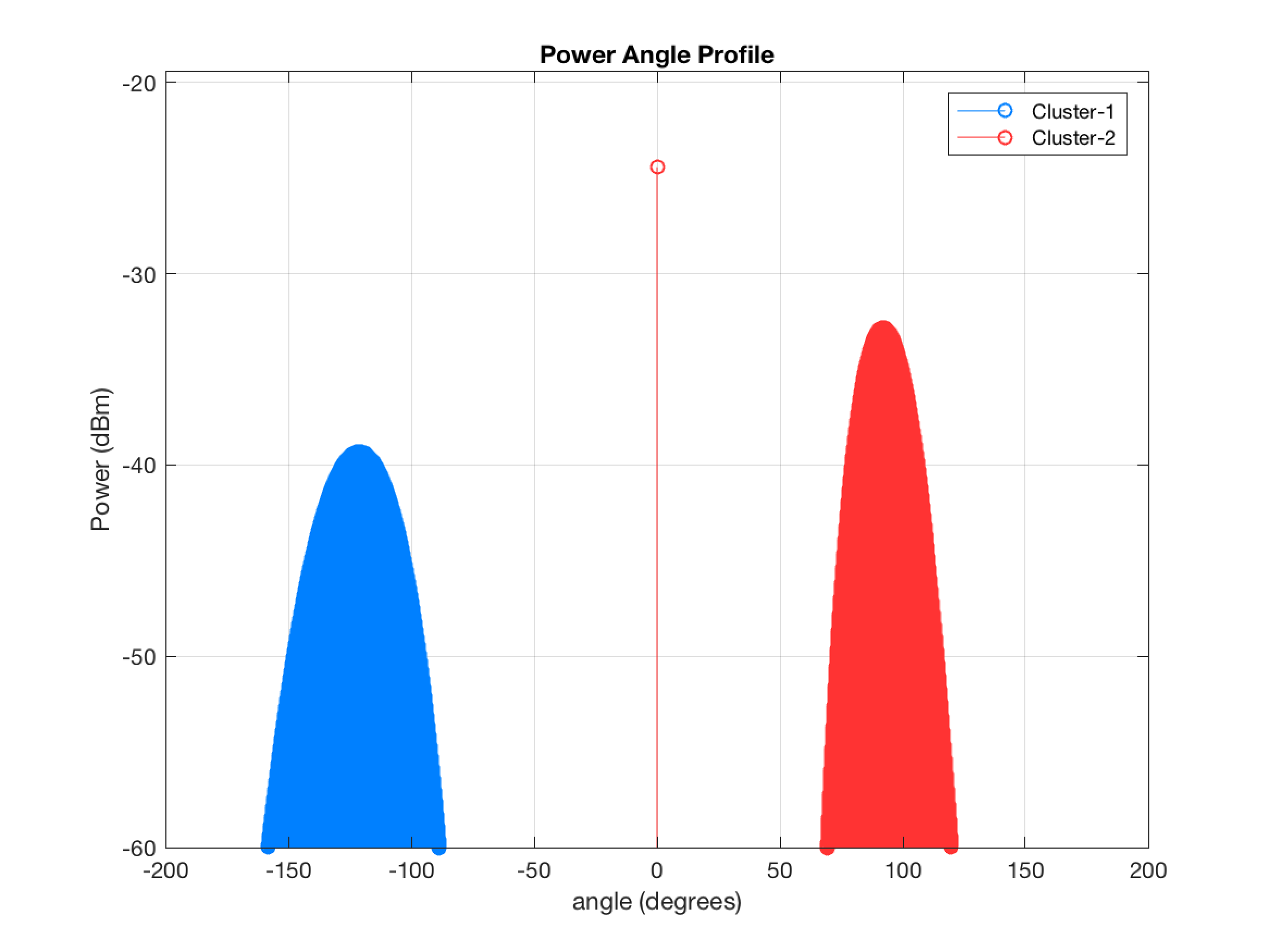} \\
\caption{Theoretical Power Angle Profile of the channel when receiver is in the corner of the room.}\label{cornertheoangle}
\end{figure}

\begin{figure}[t]
\centering
\includegraphics[scale=0.55]{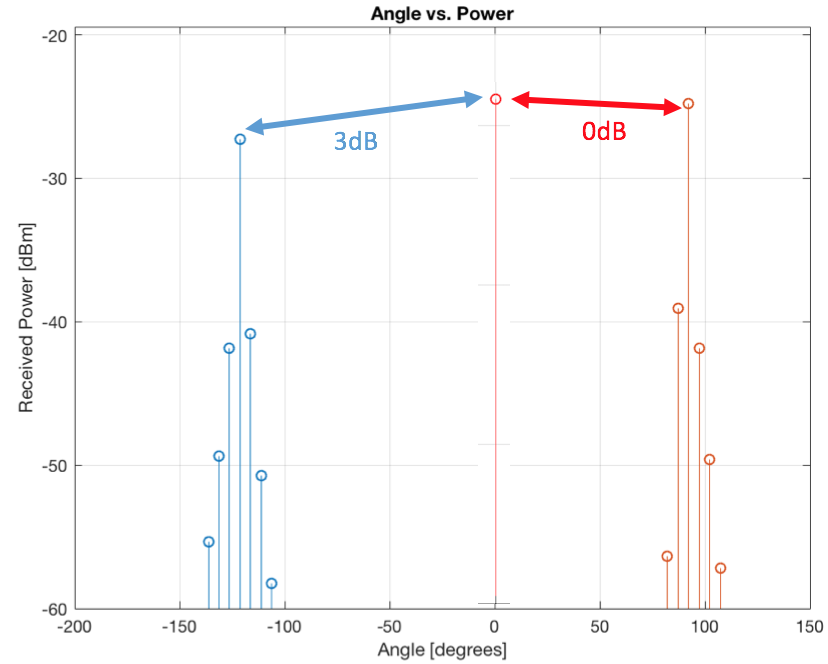} \\
\caption{Binned Power Angle Profile of the channel when receiver is in the corner of the room.}\label{cornertheoanglebin}
\end{figure}

Theoretical angle response given in \ref{cornertheoangle} shows that the cluster-1 has an AoA of $-120^{\circ}$ whereas the cluster-2 AoA is $91^{\circ}$. Their angle spreads are $69^{\circ}$ and $51^{\circ}$. The measurement result figure in \cite{Xu} gives the AoAs as $-120^{\circ}$ and $90^{\circ}$ and the spreads approximately are $50^{\circ}$ and $58^{\circ}$ for the cluster-1 and cluster-2, respectively. Similar to the previous measurement, binned version of the angle spectrum is given in Fig. \ref{cornertheoanglebin}. As seen, the relative received powers of cluster-1 and cluster-2 are $3$ dB and $0$ dB. They are approximately $7$ dB and $0$ dB in the measurements.

\section{Conclusion} \label{Sec_Conclusion}

In this paper, we create a ray tracing channel model, RT-ICM, for a mmWave channel cluster that includes only the first-order reflection rays. We also take diffuse scattering into account as the scattering has a non-negligible contribution in mmWave channels. Specifically, we aim at a spatial representation of the cluster at the receiver end. Further, since the mmWave channels are sparse and clusters are spatially separated most of the time, we claim that the proposed intra-cluster model can be generalized to the MIMO channel model simply by replicating it for each cluster. We discuss that, in fact, the transmit beamwidth can be the dominant limitation factor on the clusters angle spread in MIMO and massive MIMO applications; thereby increasing the number of first-order clusters further. After implementing the model to a literature measurement scenario, we show that the intra-cluster model estimates the angular spectrum with high accuracy.

\appendices %to add different sections, make this \appendices
\section{BGM Parameters} \label{appen_calculate}
In this Appendix, we give the complete procedure of how the BGM parameters derived. 
\subsection{Derivation of $\phi$ and $l_{dif}$} 
From Fig. \ref{basicmodel}, $\phi=\cos^{-1} (h_r/d_1)$ where $d_1=(h_r l_{sp}/(h_t+h_r))$. Plugging $d_1$, we get $\phi$.

From Fig. \ref{basicmodel}, $l_1=h_r/\cos (\phi-\alpha)$ and $l_2=\sqrt{h_t^2+(s_1^\prime)^2}$ where 
$s_1^\prime=s-s_2^\prime$ and $s_2^\prime=l_1\sin (\phi-\alpha)$. Plugging everything to Eq. (\ref{ldif}), Eq. (\ref{ldif_main}) is obtained.

\subsection{Derivation of Support Region Limitations} \label{appen_support}
\subsubsection{Geometry Limitation}
%In Figure \ref{twocases}a, $\alpha<0$, and $\beta=\phi-\alpha>0$ . According to the geometry, while $\beta$ increases the ray length increases too and once $\beta=90^{\circ}$, there doesn't exist a reflection anymore. Hence, diffuse ray AoA is bounded as $\beta<90^{\circ}$ for a valid reflection. Similarly, from Figure \ref{twocases}b, $\alpha>0$, and $\beta=\phi-\alpha<0$ when $\alpha>\phi$. And the limit for $\beta=-90^{\circ}$, which sets the bound to $\beta>-90^{\circ}$. Combining the both sides, we get Eq. (\ref{bygeo}).

From Fig. \ref{basicmodel}, for $\alpha<0$, the tilt angle doesn't increase the support region as any ray captured by the receiver with AoA of $\phi-\alpha\geq 90^{\circ}$ cannot be a reflection from that reflector. Hence, the lower bound for $\alpha$ is $\phi-90^{\circ}$. On the other hand, upper bound is a little tricky. The line goes through the transmitter and receiver (LOS line) sets the new limit to the upper bound and the tilting reduces the upper bound by $\sigma$. %This is because when the departing ray from the transmitter fits to that line, it becomes an LOS ray and %In fact, the reflector itself determines the geometric limits and the rays ahead of LOS line have still a reflection from the wall. However, since that reflection occurs behind the receiver (instead of between receiver and reflector), the distance and length calculations of the model that we formulated in Section \ref{subsub_Calc} fails to compute the parameters. Hence, we assume 
%the tilting reduces the upper bound by $\sigma$. %Note that we let $\sigma$ to be a positive angle here. 
%From Figure \ref{case2}, 
Similarly, for the case $h_t<h_r$, $\sigma$ limits $\alpha$ on the lower bound to be $\alpha> \phi-\sigma-90^{\circ}$, while the upper bound remains unchanged. %Note, since the $\sigma$ is the angle from the LOS line to the reflector orientation, in this case, $\sigma<0$. Hence, the formula for $\sigma$ in Eq. (\ref{sigma}) remains same.

\subsubsection{Reflector Length Limitation}
For $\alpha_{pos}$, $s_2=d_1\sin \phi$ %\label{s2} 
and $s_{pos}=s_2-l_{pos}$ %\label{spos}
Then, $\phi-\alpha_{pos}=\tan^{-1}s_{pos}/h_r$. 
%So, $\alpha_{pos}=\phi-\tan^{-1} \left( \frac{d_1\sin \phi-l_{pos}}{h_r} \right)$ %\label{alpha_arti}
Similarly for $\alpha_{neg}$, $\phi-\alpha_{neg}=\tan^{-1}\left( (s_2+l_{neg})/h_r\right)$.
%\begin{equation}
%\alpha_{neg}=\phi-\tan^{-1}\left( \frac{d_1\sin \phi+l_{neg}}{h_r}\right) \label{alpha_eksi}
%\end{equation}

\subsubsection{Transmitter Beamwidth Limitation}
In Fig. \ref{txbw}, from the right triangle similarity, the angle between the RNT and the departing specular ray is equal to $\phi$. %We let it to be a positive angle for ease of calculation. Similarly, both beamwidth angles are assumed to be positive and denoted with clockwise direction rows. To proceed, transmitter beamwidth has to be given to the model as an input. 
Hence, % from trigonometry,
$s_t=h_t \tan \left( \phi-(\Theta/2) \right)$. %\label{st}
On the other hand, $s_1=h_t \tan \phi$. Then $l_t=s_1-s_t$. %Plugging $s_1$ and $s_t$,
%\begin{equation}
%l_t=h_t \left( \tan (\phi) - \tan \left( \phi-\frac{\Theta}{2} \right) \right)
%\end{equation}
And for $l_r$, $s_1+l_r= h_t \tan \left( \phi+(\Theta/2) \right)$.
%l_r &=h_t \tan \left( \phi+\frac{\Theta}{2} \right)-s_1 %\numberthis \label{lr}
%\end{align*}

%Plugging $s_1$,
%\begin{equation}
%l_r =h_t \left( \tan \left( \phi+\frac{\Theta}{2} \right)- \tan \phi \right) \label{lr}
%\end{equation}

\subsection{Formulation Validation of BGM} \label{appen_formulvalid}
\subsubsection{Path Length Calculation Check} \label{SSS_pathlenCheck}

%During the formulation setup of the Basic Geometric Model throughout Section ‎\ref{subsub_Calc}, we form all the expressions according to an implicit constraint that all diffuse rays are reflecting within the reflector normals at the transmitter and the receiver. We claim that the outputs of the given expressions are valid for “outside” reflections too although some intermediate parameters don't make sense.

%First, notice that the specular ray (for both incidence and reflection portion) always stays within the reflector normals frame. Hence, parameters regarding the specular ray don't change. That is, $\l_{sp}$, $s$, $\phi$, $d_1$, $s_1$ and $s_2$ in Eq. (\ref{lspec})-(\ref{phi_latest}), Eq. (\ref{s2}) and Eq. (\ref{s1}), remain all the same. For the other equations regarding the diffuse ray, we first give the diagrams for two cases in Figure \ref{outcases} in order to explain the problem clearer.

%The equations that we seek to check are Eq. (\ref{l1}) through (\ref{s2prime}) for positive and negative side reflections given in Figure \ref{outcases}. %a and Figure \ref{outcases}b separately. %For convenience, these equations are copied below.
%\begin{equation*}
%l_1=\frac{h_r}{\cos (\phi-\alpha)}
%\end{equation*}
%\begin{equation*}
%l_2=\sqrt{h_t^2+(s_1^\prime)^2}
%\end{equation*}
%\begin{equation*}
%s_1^\prime=s-s_2^\prime
%\end{equation*}
%\begin{equation*}
%s_2^\prime=l_1\sin (\phi-\alpha) 
%\end{equation*}

%In Figure \ref{outcases}a, 
For positive side reflection, $\phi-\alpha_p<0$, but $\cos (\phi-\alpha_p)>0$, hence $l_1>0$. However, since $\sin (\phi-\alpha_p)<0$, $s_{2,p}^{\prime}<0$. Thus, $s_{1,p}^{\prime}$ is larger than $s$ but $l_2$ is accurately computed based on the geometry. As a result, %except $s_{2,p}^{\prime}$, all equations are hold. And the 
resultant calculations of $l_1$ and $l_2$ are correct.

%For $\beta>0$ shown in Figure \ref{outcases}b, 
For negative side reflection, $\phi-\alpha_n>0$, and $l_1$ is calculated as expected. However, since $s_{2,n}^{\prime}>s$, $s_{1,n}^{\prime}$ is negative. Note that, when calculating $l_2$, $s_{1,n}^{\prime}$ is squared. Hence, $l_2$ is resulted as expected too. 

%As a summary, although some intermediate variables result in negative values, the resultant output of ray lengths are always accurate. %However, care has to be taken if $s_1^{\prime}$ or $s_2^{\prime}$ is re-used in the future calculations.

\subsubsection{Reflector Length Calculation Check} \label{SSS_reflenCheck}

%Similar to the path length calculation, we formulate the reflector length limitation in Section ‎\ref{SSS_limitRef} with an implicit assumption that the reflector, itself, fits within the reflector normals frame as seen in Figure \ref{walllength}. And the formulation setup from Eq. (\ref{s2}) through (\ref{alpha_eksi}) is completed according to this assumption. We give the proof that the computation of the limitation range for $\alpha$ is accurate for the case where the reflector exceeds the reflector normal frame. The illustration of the scenario is given in Figure \ref{outreflen}.

%We check the formulation setup from Eq. (\ref{spos}) through (\ref{alpha_eksi}). 
%Notice that Eq. (\ref{s2}) remains same since the specular ray parameters don't change as discussed in previous subsection. Hence, we check the rest of the equations. %For convenience again, they are copied below.
%\begin{equation*}
%s_{pos}=s_2-l_{pos}
%\end{equation*}
%\begin{equation*}
%\alpha^+=\phi-\tan^{-1} \left( \frac{d_1\sin \phi-l_{pos}}{h_r} \right) 
%\end{equation*}
%\begin{equation*}
%\alpha^-=\phi-\tan^{-1}\left( \frac{d_1\sin \phi+l_{neg}}{h_r}\right) 
%\end{equation*}

For positive side reflection, since $l_{pos}$ is larger than $s_2$, $s_{pos}$ turns out to be negative. That yields $\phi-\alpha_{pos}<0$ which is, actually, correct as $\alpha_{pos}$ is larger than $\phi$.  %That is, the result is accurate for $\beta<0$. 
For negative side reflection, nothing is unusual in the formulation.% although $s_{neg}<0$ for this case. However, we even don't use it in our calculations. In anyway, we give the notation for it for a possible future reference.
%As a summary, formulation that computes the limitation of the reflector length on $\alpha$ works for all scenarios too.

\subsubsection{Transmit Beamwidth Calculation Check}\label{SSS_TBWCheck}

%Finally, we check the equations for transmitter beamwidth limitation of the reflector length that are given from Eq. (\ref{st}) to (\ref{lr}) in case of beamwidth incident rays exceed the reflector normal frame as shown in Figure \ref{outtxbw}.

%The equations that are checked are copied below, again. 
%\begin{equation*}
%s_t=h_t \tan \left( \phi-\frac{\Theta}{2} \right)
%\end{equation*}
%\begin{equation*}
%l_t=s_1-s_t
%\end{equation*}
%\begin{equation*}
%l_r =h_t \tan \left( \phi+\frac{\Theta}{2} \right)-s_1
%\end{equation*}

As seen from Fig. \ref{txbw}, $\phi-\Theta/2<0$ which yields $s_t<0$. However, $l_t$ is calculated correctly. For $l_r$, calculation is as expected. 
%As a result, reflector length limitation due to transmitter beamwidth is calculated appropriately for all possible scenarios. This completes the proof of the claim that out of frame scenarios don't change any output parameter of the model.

\section{Validation of the Directive Model for All Cases} \label{appen_diffuseValidation}

%Eq. (\ref{thetak}) and (\ref{psik}) are formed based on the diagram given in Figure \ref{diffuseupdate}. In that scenario, a diffuse ray reflected from the transmitter side of the specular ray is considered. There are other cases need to be checked in order to ensure that the equations hold for any scenario.
We consider the cases, diffuse rays reflected from (1) the \textit{receiver} side of the specular ray, (2) the back of the RNR, (3) the back of the RNT. For the case (1), the diffuse rays in Fig. \ref{diffusemod} with $\theta_1$ can be an example. In that case, $\alpha_k$ and $\psi_k$ are positive. Eq. (\ref{thetak}) holds as the variables don't change. Since we paid attention to the angle signs during the formulation setup, Eq. (\ref{psik}) holds too. For the case (2), the diffuse ray in Fig. \ref{diffusemod} with $\theta_2$ is an example. $\alpha_k$ and $\psi_k$ are still positive. From Fig. \ref{outcases}, $s_1^{\prime}$ is positive and grazing angle calculation in Eq. (\ref{thetak}) is valid. Since $\alpha_k>\phi$, $(\phi-\alpha_k)<0$. Hence, $\theta_k+\psi_k>90$ which is the case as spread angle exceeds the reflector normal at reflection point. Hence, Eq. (\ref{psik}) is valid too. However, in the case of (3), for which the diffuse ray in Fig. \ref{diffusemod} with $\theta_3$ is an example, the specular reflection of the diffuse ray reflects towards the opposite direction of the receiver. However, Eq. (\ref{psik}) computes $\psi_k^{\prime}$ as shown in Fig. \ref{diffusemod} which is inaccurate. To correct it, additional $2(90-\theta_k)$ should be added. That is, recalling that $\psi_k<0$, $\psi_k=90-(\phi-\alpha_k)-\theta_k-2(90-\theta_k)=\theta_k-90-(\phi-\alpha_k)$.

\section*{Acknowledgement}
The authors would like to thank Prof. S. Orfanidis for many helpful discussions and his contributions to the effect of reflection and scattering propagation mechanisms as well as the concepts regarding the antenna theory. 

\bibliographystyle{IEEEtran}

\begin{thebibliography}{1}
\footnotesize

%mmWave general
\bibitem{Rangan}
Rangan, Sundeep, Theodore S. Rappaport, and Elza Erkip. "Millimeter-wave cellular wireless networks: Potentials and challenges." \textit{Proceedings of the IEEE} 102.3 (2014): 366-385.

\bibitem{Yaman}
Yaman, Yavuz, and Predrag Spasojevic. "Reducing the LOS ray beamforming setup time for IEEE 802.11ad and IEEE 802.15.3c." \textit{Military Communications Conference, MILCOM 2016-2016 IEEE.} IEEE, 2016.

\bibitem{Ayach}
El Ayach, Omar, et al. "Spatially sparse precoding in millimeter wave MIMO systems." \textit{IEEE transactions on wireless communications}, 13.3 (2014): 1499-1513.

%\bibitem{Bjornson} ---> check
%Bjornson, Emil, et al. "Massive MIMO in Sub-6 GHz and mmWave: Physical, Practical, and Use-Case Differences." \textit{arXiv preprint arXiv}:1803.11023 (2018).

%\bibitem{Bogale} ---> check
%Bogale, Tadilo Endeshaw, and Long Bao Le. "Massive MIMO and mmWave for 5G wireless HetNet: Potential benefits and challenges." \textit{IEEE Vehicular Technology Magazine} 11.1 (2016): 64-75.

%\bibitem{Swindlehurst} ---> check
%Swindlehurst, A. Lee, et al. "Millimeter-wave massive MIMO: The next wireless revolution?." \textit{IEEE Communications Magazine} 52.9 (2014): 56-62.

%\bibitem{Sun} ---> check
%Sun, Shu, et al. "MIMO for millimeter-wave wireless communications: Beamforming, spatial multiplexing, or both?." \textit{IEEE Communications Magazine} 52.12 (2014): 110-121.

%Channel Modelling References
%-Journals
%\bibitem{Ertel} --->check
%Ertel, Richard Brian, et al. "Overview of spatial channel models for antenna array communication systems." \textit{IEEE personal communications} 5.1 (1998): 10-22.

\bibitem{3GPP} 
 "Technical Specification Group Radio Access Network; Study on Channel Model for Frequencies from 0.5 to 100 GHz (Rel. 14)", \textit{3GPP TR 38.901 V14.1.1 (2017–07),} July 2017.
 
\bibitem{MiWEBA}
Maltsev, A. "D5. 1-Channel Modeling and Characterization." \textit{MiWEBA Project (FP7-ICT-608637), Public Deliverable} (2014).

\bibitem{COST2100}
Liu, Lingfeng, et al. "The COST 2100 MIMO channel model." \textit{IEEE Wireless Communications} 19.6 (2012): 92-99.

\bibitem{Gustafson}
Gustafson, Carl. \textit{60 GHz Wireless Propagation Channels: Characterization, Modeling and Evaluation}. Vol. 69. 2014.

\bibitem{11ad}
Maltsev A., et al, “Channel models for 60 GHz WLAN systems,” \textit{IEEE Document} 802.11-09/0334r6, January 2010.

\bibitem{11ay}
https://mentor.ieee.org/802.11/dcn/15/11-15-1150-09-00ay-channel-models-for-ieee-802-11ay.docx

\bibitem{3c}
S-K. Yong, et al., TG3c channel modeling sub-commitee final report, \textit{IEEE Techn. Rep.},15-07-0584-01-003c, Mar. 2007.

\bibitem{Saleh}
Saleh, Adel AM, and Reinaldo Valenzuela. "A statistical model for indoor multipath propagation." \textit{IEEE Journal on selected areas in communications} 5.2 (1987): 128-137.

\bibitem{Spencer}
Spencer, Quentin H., et al. "Modeling the statistical time and angle of arrival characteristics of an indoor multipath channel." \textit{IEEE Journal on Selected areas in communications} 18.3 (2000): 347-360.

\bibitem{Steinbauer}
Steinbauer, Martin, Andreas F. Molisch, and Ernst Bonek. "The double-directional radio channel." \textit{IEEE Antennas and propagation Magazine} 43.4 (2001): 51-63.

%\bibitem{Durgin_journal} %not cited. but may be in implementation sec
%+Durgin, Gregory D., and Theodore S. Rappaport. "Theory of multipath shape factors for small-scale fading wireless channels." \textit{IEEE Transactions on Antennas and Propagation} 48.5 (2000): 682-693.

\bibitem{Xu}
Xu, Hao, Vikas Kukshya, and Theodore S. Rappaport. "Spatial and temporal characteristics of 60-GHz indoor channels." \textit{IEEE Journal on selected areas in communications} 20.3 (2002): 620-630.

\bibitem{Akdeniz}
Akdeniz, Mustafa Riza, et al. "Millimeter wave channel modeling and cellular capacity evaluation." \textit{IEEE journal on selected areas in communications} 32.6 (2014): 1164-1179.

\bibitem{Maltsev_office}
Maltsev, Alexander, et al. "Experimental investigations of 60 GHz WLAN systems in office environment." \textit{IEEE Journal on Selected Areas in Communications} 27.8 (2009).

%\bibitem{Geng} %not cited. check again
%+Geng, Suiyan, et al. "Millimeter-wave propagation channel characterization for short-range wireless communications." \textit{IEEE Transactions on Vehicular Technology} 58.1 (2009): 3-13.

\bibitem{Smulders}
Smulders, Peter FM. "Statistical characterization of 60-GHz indoor radio channels." \textit{IEEE Transactions on Antennas and Propagation} 57.10 (2009): 2820-2829.

\bibitem{Kyro}
Kyro, Mikko, et al. "Statistical channel models for 60 GHz radio propagation in hospital environments." \textit{IEEE Transactions on Antennas and Propagation} 60.3 (2012): 1569-1577.

\bibitem{Samimi}
Samimi, Mathew K., and Theodore S. Rappaport. "3-D millimeter-wave statistical channel model for 5G wireless system design." \textit{IEEE Transactions on Microwave Theory and Techniques} 64.7 (2016): 2207-2225.

\bibitem{Gustafson_journal}
Gustafson, Carl, et al. "On mm-wave multipath clustering and channel modeling." \textit{IEEE Transactions on Antennas and Propagation} 62.3 (2014): 1445-1455.

%\bibitem{Rappaport_wideband} %not cited. check again
%+Rappaport, Theodore S., et al. "Wideband millimeter-wave propagation measurements and channel models for future wireless communication system design." \textit{IEEE Transactions on Communications} 63.9 (2015): 3029-3056.

\bibitem{Weiler} %not cited. check again.
Weiler, Richard J., et al. "Quasi-deterministic millimeter-wave channel models in MiWEBA." \textit{EURASIP Journal on Wireless Communications and Networking} 2016.1 (2016): 84.

%\bibitem{Rappaport_broadband} %not cited. check again.
%+Rappaport, Theodore S., et al. "Broadband millimeter-wave propagation measurements and models using adaptive-beam antennas for outdoor urban cellular communications." \textit{IEEE transactions on antennas and propagation} 61.4 (2013): 1850-1859.

\bibitem{Gentile}
Gentile, Camillo, et al. "Quasi-Deterministic Channel Model Parameters for a Data Center at 60 GHz." \textit{IEEE Antennas and Wireless Propagation Letters} 17.5 (2018): 808-812.

%Books
\bibitem{Rappaport_book}
Rappaport, Theodore S. \textit{Wireless communications: principles and practice}. Vol. 2. New Jersey: prentice hall PTR, 1996.

\bibitem{Orfanidis}
Orfanidis, Sophocles J. "Electromagnetic waves and antennas." (2002).

\bibitem{Glassner}
Glassner, Andrew S., ed. \textit{An introduction to ray tracing}. Elsevier, 1989.

%Conference Papers
%\bibitem{Durgin} ---> check
%Durgin, Greg, Neal Patwari, and Theodore S. Rappaport. "An advanced 3D ray launching method for wireless propagation prediction." \textit{Vehicular Technology Conference}, 1997, IEEE 47th. Vol. 2. IEEE, 1997.

%\bibitem{Schaubach} ---> check
%Schaubach, Kurt R., N. J. Davis, and Theodore S. Rappaport. "A ray tracing method for predicting path loss and delay spread in microcellular environments." \textit{Vehicular Technology Conference, 1992, IEEE 42nd}. IEEE, 1992.

%\bibitem{Moraitis} %not cited. check again.
%+Moraitis, Nektarios, and Philip Constantinou. "Propagation modeling at 60 GHz for indoor wireless LAN Applications." \textit{Proc. IST.} Vol. 2. 2002.

%\bibitem{Czink} ---> check
%Czink, Nicolai, et al. "A framework for automatic clustering of parametric MIMO channel data including path powers." \textit{Vehicular Technology Conference, 2006. VTC-2006 Fall. 2006 IEEE 64th.} IEEE, 2006.

%\bibitem{Moraitis_PAP} %not cited. check again. Check Excel file
%+Moraitis, Nektarios, Philip Constantinou, and Demosthenes Vouyioukas. "Power angle profile measurements and capacity evaluation of a SIMO system at 60 GHz." \textit{Personal Indoor and Mobile Radio Communications (PIMRC), 2010 IEEE 21st International Symposium on.} IEEE, 2010.

\bibitem{Maltsev_conference}
Maltsev, Alexander, et al. "Statistical channel model for 60 GHz WLAN systems in conference room environment." \textit{Antennas and Propagation (EuCAP), 2010 Proceedings of the Fourth European Conference on.} IEEE, 2010.

\bibitem{Murdock}
Murdock, James N., et al. "A 38 GHz cellular outage study for an urban outdoor campus environment." \textit{Wireless Communications and Networking Conference (WCNC), 2012 IEEE.} IEEE, 2012.

\bibitem{Rappaport_aoa}
Rappaport, Theodore S., et al. "Cellular broadband millimeter wave propagation and angle of arrival for adaptive beam steering systems." \textit{Radio and Wireless Symposium (RWS), 2012 IEEE.} IEEE, 2012.

\bibitem{Thomas}
Thomas, Timothy A., et al. "3D mmWave channel model proposal." \textit{Vehicular Technology Conference (VTC Fall), 2014 IEEE 80th.} IEEE, 2014.

%probably remove this - not new stuff
%\bibitem{Wang} 
%Wang, Lei, et al. "First-order-reflection MIMO channel model for 60 GHz NLOS indoor WLAN systems." \textit{Communication Systems (ICCS), 2014 IEEE International Conference on.} IEEE, 2014.

\bibitem{Maltsev_QD} %not cited. check some MiWEBA references.  
Maltsev, Alexander, et al. "Quasi-deterministic approach to mmwave channel modeling in a non-stationary environment." \textit{Globecom Workshops (GC Wkshps), 2014.} IEEE, 2014.

%\bibitem{Maltsev_11ay} %not cited. but must be in site-specific channel model
%+Maltsev, Alexander, et al. "Channel modeling in the next generation mmWave Wi-Fi: IEEE 802.11 ay standard." \textit{European Wireless 2016; 22th European Wireless Conference; Proceedings of.} VDE, 2016. +

%\bibitem{Wang} ---> check
%Wang, Jian, et al. "Unsupervised clustering for millimeter-wave channel propagation modeling." \textit{Vehicular Technology Conference (VTC-Fall), 2017 IEEE 86th.} IEEE, 2017.

\bibitem{Rajagopal}
Rajagopal, Sridhar, Shadi Abu-Surra, and Mehrzad Malmirchegini. "Channel feasibility for outdoor non-line-of-sight mmwave mobile communication." \textit{Vehicular Technology Conference (VTC Fall), 2012 IEEE.} IEEE, 2012.

%\bibitem{Medbo} %not cited. but must be. check again.
%+Medbo, J., et al. "Channel modelling for the fifth generation mobile communications." \textit{Antennas and Propagation (EuCAP), 2014 8th European Conference on.} IEEE, 2014.

%Diffuse Scattering References
%ITU-report
%\bibitem{Hayn} ---> check
%Hayn, A., R. Bose, and R. Jakoby. "Multipath propagation and LOS interference studies for LMDS architecture." (2001): 686-690.

%\bibitem{Hammoudeh} ---> check
%Hammoudeh, A., M. G. Sanchez, and E. Grindrod. "Modelling of propagation in outdoor microcells at 62.4 GHz." \textit{Microwave Conference, 1997. 27th European.} Vol. 1. IEEE, 1997.

\bibitem{ITU}
ITU Report 1008-1. "Reflection from the Surface of the Earth". 1986-1990.

\bibitem{ITU2}
ITU-R  P.2040-1. "Effects of building materials and structures on radiowave propagation above about 100 MHz".  (07/2015). 

%\bibitem{Ament} %not cited. 
%+Ament, W. S. "Toward a theory of reflection by a rough surface." \textit{Proceedings of the IRE} 41.1 (1953): 142-146.

%\bibitem{Langen} ---> check
%Langen, B., G. Lober, and W. Herzig. "Reflection and transmission behaviour of building materials at 60 GHz." \textit{Personal, Indoor and Mobile Radio Communications, 1994. Wireless Networks-Catching the Mobile Future., 5th IEEE International Symposium on.} Vol. 2. IEEE, 1994.

%ER Model
\bibitem{Esposti}
Degli-Esposti, Vittorio, et al. "Measurement and modelling of scattering from buildings." \textit{IEEE Transactions on Antennas and Propagation} 55.1 (2007): 143-153.

%will check this again. specular reflections make me cautious.
%\bibitem{Virk}
%Virk, Usman Tahir, Jean-Frederic Wagen, and Katsuyuki Haneda. "Simulating specular reflections for point cloud geometrical database of the environment." \textit{Antennas and Propagation Conference (LAPC), 2015 Loughborough.} IEEE, 2015.

%will check this again. I thought 3 papers for ER model is enough
%\bibitem{Lu}
%Lu, Jonathan S., et al. "Study of back-scattering from buildings using a 60 GHz scaled model." \textit{Antennas and Propagation (EuCAP), 2014 8th European Conference on.} IEEE, 2014.

%not mmWave
%\bibitem{Minghini}
%Minghini, Lorenzo, et al. "Electromagnetic simulation and measurement of diffuse scattering from building walls." \textit{Antennas and Propagation (EuCAP), 2014 8th European Conference on}. IEEE, 2014.

%not mmWave
%\bibitem{Vitucci}
%Vitucci, Enrico M., Franco Fuschini, and Vittorio Degli-Esposti. "Ray Tracing simulation of the radio channel time-and angle-dispersion in large indoor environments." \textit{Antennas and Propagation (EuCAP), 2014 8th European Conference on.} IEEE, 2014.

% not mmW
%\bibitem{Oestges}
%Oestges, Claude, et al. "A geometry-based physical-statistical model of land mobile satellite channels in urban environments." \textit{Antennas and Propagation (EuCAP), 2014 8th European Conference on.} IEEE, 2014.

%\bibitem{Esposti_rayTracing} ---> check
%Degli-Esposti, Vittorio, et al. "Ray-tracing-based mm-wave beamforming assessment." \textit{IEEE Access 2} (2014): 1314-1325.

%this paper is completed in Pascual3
%\bibitem{Pascual}
%Pascual-Garcia, Juan, et al. "Using tuned diffuse scattering parameters in ray tracing channel modeling." \textit{Antennas and Propagation (EuCAP), 2015 9th European Conference on.} IEEE, 2015.

%\bibitem{Pascual2} ---> check
%Pascual-Garcia, Juan, et al. "Experimental parameterization of a diffuse scattering model at 60 GHz." \textit{Antennas and Propagation in Wireless Communications (APWC), 2015 IEEE-APS Topical Conference on}. IEEE, 2015.

\bibitem{Pascual3}
Pascual-García, Juan, et al. "On the importance of diffuse scattering model parameterization in indoor wireless channels at mm-wave frequencies." \textit{IEEE Access 4} (2016): 688-701.

%not mmWave
%\bibitem{Wagen}
%Wagen, Jean-Frederic, Usman Tahir Virk, and Katsuyuki Haneda. "Measurements based specular reflection formulation for point cloud modelling." \textit{Antennas and Propagation (EuCAP), 2016 10th European Conference on.} IEEE, 2016.

%not mmWave
%\bibitem{Tran}
%Tran, Ngochao, et al. "A study on wall scattering characteristics based on ER model with point cloud data." \textit{Computational Electromagnetics (ICCEM), 2017 IEEE International Conference on}. IEEE, 2017.

%\bibitem{Inomata} ---> check
%Inomata, Minoru, et al. "Prediction accuracy of hybrid method based on ray-tracing and effective roughness model in indoor environment for millimeter waves." \textit{Antenna Measurements and Applications (CAMA), 2017 IEEE Conference on.} IEEE, 2017.

%Other Diffuse Scattering
%\bibitem{Sarabandi} --->check
%Sarabandi, Kamal, Eric S. Li, and Abid Nashashibi. "Modeling and measurements of scattering from road surfaces at millimeter-wave frequencies." \textit{IEEE Transactions on Antennas and Propagation} 45.11 (1997): 1679-1688.

%already cited main paper
%\bibitem{Li}
%Li, Eric S., and Kamal Sarabandi. "Low grazing incidence millimeter-wave scattering models and measurements for various road surfaces." \textit{IEEE Transactions on Antennas and Propagation} 47.5 (1999): 851-861.

\end{thebibliography}

%Authors
\newpage
\begin{IEEEbiographynophoto}{Yavuz Yaman}
received the B.S degree from the School of Engineering, Istanbul University, in 2011; M.S. degree in electrical and computer engineering from Rutgers University, Piscataway, NJ, in 2014. He is currently working toward the Ph.D. degree with the Department of Electrical and Computer Engineering, Rutgers University, Piscataway, NJ. His research interests include channel modelling, beamforming, channel estimation, antenna propagations and phased antenna arrays.
\end{IEEEbiographynophoto}

\begin{IEEEbiographynophoto}{Predrag Spasojevic}
received the Diploma of Engineering degree from
the School of Electrical Engineering, University of Sarajevo, in 1990;
and M.S. and Ph.D. degrees in electrical engineering from Texas A\&M
University, College Station, Texas, in 1992 and 1999, respectively.
From 2000 to 2001, he was with WINLAB, Electrical and Computer
Engineering Department, Rutgers University, Piscataway, NJ, as a
Lucent Postdoctoral Fellow. He is currently  Associate Professor in
the Department of Electrical and Computer Engineering, Rutgers
University. Since 2001 he is a member of the WINLAB research faculty.
 From 2017 to 2018 he was a Senior Research Fellow with
Oak Ridge Associated Universities working at the Army Research Lab, Adelphi, MD.
 His research interests are in the general areas of
communication and information theory, and signal processing.

Dr. Spasojevic was an Associate Editor of the IEEE Communications
Letters from 2002 to 2004 and served as a co-chair of the DIMACS
Princeton-Rutgers Seminar Series in Information Sciences and Systems
2003-2004. He served as a Technical Program Co-Chair for IEEE Radio
and Wireless Symposium in 2010. From 2008-2011 Predrag served as the
Publications Editor of IEEE Transactions of Information Theory.
\end{IEEEbiographynophoto}

%\newpage


\end{document}